\documentclass[%
superscriptaddress,
twocolumn,
showkeys,
nofootinbib,
nobibnotes,
 amsmath,amssymb,
pra
]{revtex4-1}
\usepackage[utf8]{inputenc}
\DeclareUnicodeCharacter{2010}{-}
\usepackage{multirow}
\usepackage[caption=false]{subfig}
\usepackage{graphicx}
\usepackage{nicefrac}
\usepackage{todonotes}
\usepackage[colorlinks,
pdfpagelabels,
pdfstartview = FitH,
bookmarksopen = true,
bookmarksnumbered = true,
linkcolor = black,
plainpages = false,
hypertexnames = false,
citecolor = black] {hyperref}

\usepackage[version=3]{mhchem}
\usepackage{color}
\usepackage[range-phrase = --,range-units = single,exponent-product = \cdot]{siunitx}
\DeclareSIUnit\atom{atom}
\usepackage{dcolumn}
\newcolumntype{d}[1]{D{.}{}{#1}}
\newcolumntype{k}[1]{D{.}{.}{#1}}
\usepackage{xcolor,colortbl}
\usepackage{tikz}

\newcommand\tikzmark[2]{%
  \tikz[remember picture]\node[inner sep=0pt,outer sep=0pt] (#1) {#2};%
}

\newcommand\ConnectC[3][]{%
\tikz[remember picture,overlay]%
  \draw[line width=10pt,-stealth,black,opacity=0.2] ([xshift=-10mm]#2) -- ([xshift=-10mm]#3);
}
\newcommand\ConnectBN[3][]{%
\tikz[remember picture,overlay]%
  \draw[line width=10pt,-stealth,green,opacity=0.2] ([xshift=10mm]#2) -- ([xshift=+10mm]#3);
}
\newcommand\ConnectD[4][]{%
\tikz[remember picture,overlay]%
  \draw[line width=10pt,-stealth,#4,opacity=0.2] (#2) -- (#3);
}

\newcommand\Connectbow[7][]{%
\tikz[remember picture,overlay]%
  \draw[draw,line width=10pt,-stealth,#6,opacity=0.2,bend angle=#4] ([yshift=0mm]#2) to [distance=#7, bend #5] ([xshift=-0mm]#3);
}

\begin{document}

\newcommand{\GABC}{G-ABC}
\newcommand{\GAB}{G-AB}
\newcommand{\GAA}{G-AA}
\newcommand{\cD}{\text{$c$-D}}
\newcommand{\hD}{\text{$h$-D}}
\newcommand{\cBN}{\text{$c$-BN}}
\newcommand{\wBN}{\text{$w$-BN}}
\newcommand{\hBN}{\text{$h$-BN}}
\newcommand{\rBN}{\text{$r$-BN}}
\newcommand{\BNAB}{\text{BN-AB}}
\newcommand{\bwTS}{\text{bw-TS}}
\newcommand{\pwTS}{\text{pw-TS}}
\newcommand{\pcTS}{\text{pc-TS}}
\newcommand{\lpcTS}{\text{l-pc-TS}}

\newcommand{\fref}[1]{Fig.~\ref{#1}}
\newcommand{\fsref}[1]{Fig.~\ref{#1}~(Supporting Information)}
\newcommand{\tref}[1]{Tab.~\ref{#1}}
\newcommand{\tsref}[1]{Table~\ref{#1}~(Supporting Information)}
\newcommand{\eref}[1]{Eq.~\ref{#1}}
\newcommand{\rcite}[1]{Ref.~\ref{#1}}
\newcommand{\abinitio}{\emph{ab initio}}
\newcommand{\etal}{et al.}
\newcommand{\rcm}{\ensuremath{\mathrm{cm}^{-1}}}
\newcommand{\lt}[1]{\ensuremath{_\mathrm{#1}}}
\newcommand{\ut}[1]{\ensuremath{^\mathrm{#1}}}
\def\figw{0.21\textwidth}
\def\figs{\qquad}

\makeatletter
\renewcommand\@dotsep{10000}
\makeatother


\title[]{$Ab-initio$ calculations of carbon and boron nitride allotropes
and their structural phase transitions using periodic coupled cluster theory}
\date{\today}
\author{Thomas Gruber} 
\affiliation{Max-Planck-Institute for Solid State Research, Heisenbergstraße 1, 70569 Stuttgart, Germany}
\author{Andreas Grüneis}
\email{andreas.grueneis@tuwien.ac.at}
\affiliation{Max-Planck-Institute for Solid State Research, Heisenbergstraße 1, 70569 Stuttgart, Germany}
\affiliation{Institute for Theoretical Physics, Vienna University of Technology, Wiedner Hauptstrasse 8-10, 1040 Vienna, Austria}


\begin{abstract}
We present an $ab-initio$ study of boron nitride as well as carbon allotropes.
Their relative thermodynamic stabilities and structural phase
transitions from low- to high-density phases are investigated.
Pressure-temperature phase diagrams
are calculated and compared to experimental findings.
The calculations are performed using quantum chemical wavefunction based 
as well as density functional theories.
Our findings reveal that predicted energy differences often depend significantly
on the choice of the employed method.
Comparison between calculated and experimental results allows for benchmarking
the accuracy of various levels of theory.
The produced results show that quantum chemical wavefunction based theories allow for achieving systematically improvable
estimates.
We find that on the level of coupled cluster theories the low- and high-density phases of
boron nitride become thermodynamically degenerate at 0~K. This is in agreement with recent
experimental findings, indicating that cubic boron nitride is not the thermodynamically stable allotrope at
ambient conditions.
Furthermore we employ the calculated results to assess transition probabilities from graphitic low-density
to diamond-like high-density phases in an approximate manner.
We conclude that the stacking order of the parent graphitic material is crucial
for the possible formation of meta-stable wurtzite boron nitride and hexagonal carbon diamond also known as lonsdaleite.
\end{abstract}

\maketitle

\section{\label{s:intro}Introduction}

The pressure-temperature phase diagrams of carbon and boron nitride reflect a delicate balance
between weak and strong interatomic interactions.
Although covalent bonds are the main source of their large cohesive energies,
the accumulation of weak van der Waals interactions contributes significantly
to the relative stability of their low- and high-density phases.
Furthermore vibrational effects play a crucial role in
the temperature dependence of the equilibrium phase boundary.
Altogether this makes the prediction of phase diagrams
and structural phase transition pathways a challenging task for modern electronic structure theories.
In this work we seek to investigate boron nitride as well as carbon allotropes using
various approximate electronic structure theories and compare theoretical with experimental findings.
The aim is to benchmark their accuracy and help interpreting experimental results better if possible.
To this end we employ a range of approximate density functional theories (DFT)
and quantum chemical wavefunction based methods.

During the last decades approximate exchange and correlation (XC) density functionals have made significant progress
in becoming more accurate and predictive for the description of interatomic interactions while keeping a high level
of computational efficiency that allows for studying systems containing several hundreds of atoms
routinely.
The so-called Jacob's ladder describes a ladder of approximations for the XC energy
using increasingly complex as well as in general more accurate methods~\cite{doi:10.1063/1.1390175}.
These rungs include functionals based on
the local density approximation (LDA)~\cite{PhysRevLett.45.566,PhysRevB.23.5048},
the generalized gradient approximation (GGA)~\cite{PhysRevLett.77.3865}, the meta generalized gradient
approximation (mGGA)~\cite{PhysRevLett.115.036402} and hybrid functionals~\cite{doi:10.1063/1.472933,doi:10.1063/1.464913,doi:10.1063/1.1564060,doi:10.1063/1.2404663}.
The latter include a fraction of (screened) exact exchange energies
and have a computational cost that is comparable to Hartree--Fock theory.
However, all the functionals mentioned above suffer from shortcomings that are despite many efforts difficult
to remedy~\cite{doi:10.1021/cr200107z}.
In the context of the present work a significant shortcoming is the inaccurate description of long range van
der Waals interactions.
In order to describe van der Waals and related interatomic interactions more accurately in the framework of
approximate XC density functionals, a wide variety of dispersion corrections has been developed.
As a consequence of the large number of available density functionals and corrections, there are
numerous ground state energy functionals that could be considered in the present work, of which
we have only chosen a small selection.

As a complement to the treatment of exchange and correlation on the level approximate density functionals,
the computationally significantly more expensive quantum chemical wavefunction based theories are becoming
more popular for the study of periodic systems~\cite{Booth2013,Yang640,Muller2012,GrueneisJoCTaC2011,
Usvyat2013b,Nolan2009,Hirata2004a,Rosciszewski1999-me,Stoll2012,Ren2012,Del_Ben2013-or,Neuhauser2013,
doi:10.1063/1.5003794,Hermann2008,Schwerdtfeger2010-ir,McClain2017,JCC:JCC24462,Ochi2015a,
doi:10.1002/wcms.1357}.
This can partly be attributed to the increase in their computational efficiency, due to methodological developments,
and to their ability to predict exchange and correlation energies in a systematically improvable manner.
Quantum chemical methods constitute a hierarchy, which starting from the one-particle Hartree–Fock (HF) approximation,
allows for a systematic treatment of the quantum many-body effects. The simplest form of such correlated methods is
the second-order M\o{}ller–Plesset (MP2) perturbation theory~\cite{PhysRev.46.618}.
The next level of theory that achieves a significantly improved trade-off between accuracy and
computational cost is based on the coupled cluster ansatz for the many-electron wavefunction~\cite{doi:10.1063/1.1727484}.
Coupled cluster singles and doubles theory provides a compelling framework of infinite-order approximations
in the form of an exponential of cluster operators~\cite{Bartlett2007-cm}.
The coupled-cluster singles and doubles (CCSD) method where the triples are treated in a perturbative way,
termed as CCSD(T), achieves chemical accuracy in the description of many molecular properties and is sometimes
referred to as the gold standard method~\cite{RAGHAVACHARI1989479}.

In this work we seek to investigate the accuracy of the electronic structure theories mentioned above
for carbon and boron nitride allotropes.
To this end we compare predicted ground state energy differences as well as calculated phase diagrams to experimental findings.
Experimentally several phases have been synthesized as single crystals 
\cite{KubotaCoM2008,LuNC2015,TaniguchiJoCG2001, AustermanC1967,KandaBJoP2000} or as powder \cite{NagakuboAPL2013,MatsuiJoMS1981}. 
Single crystals are usually synthesized in a closed chamber over a longer time period crystallizing from a solution. 
For synthesizing metastable structures the samples are put under static pressure and heated electrically or by laser \cite{EndoPRB1994,YagiPRB1992,BritunJoMS1993,KurdyumovDaRM1996,TaniguchiDaRM1997}.
Another way of transforming samples into metastable phases is shock wave synthesis \cite{WheelerMRB1975,SatoJotACS1982}. 
With these materials thermodynamic characterization can be performed \cite{WagmanJoRotNBoS1945,Madelung2002,SolozhenkoTA1993,SolozhenkoHPR1995,JeongJoNaN2013,WiseTJoPC1966}
by determining relative enthalpy, entropy and heat capacity. 
These properties can be used to compare with the calculated energy differences and construct phase diagrams. 
Furthermore the phase diagrams can also be obtained by observing phase transitions directly 
\cite{BundyC1996,ClarkeAiP1984,CorriganTJoCP1975,EremetsPRB1998,OnoderaTJoCP1981,SachdevDaRM1997,WillsIJoHTC1985,FukunagaDaRM2000}.
In the present work we also investigate pressure-driven concerted phase transition pathways.
In particular we study activation barrier heights for the transformation from low-density to 
high-density systems considering  small unit cells that contain a few atoms at most.
These models are far from realistic conditions under
which phase transitions occur in experiment.
Temperature-driven kinetic effects and catalysts are needed in practice to observe phase transitions close to the
equilibrium phase boundary \cite{BermanZfEBdBfpC1955,BundyTJoCP1961,BundyJoGRSE1980,BundyTJoCP1963}.
Hot liquid metals can be used to dissolve graphite and diamond will precipitate at the cooler region.
However, the aim of the current work is to explore the accuracy of various electronic structure theories
for transition states occurring in these phase transitions and provide a qualitative description
of the various possible phase transition mechanisms. We believe that the small supercells considered
are sufficient to describe these effects qualitatively correct.

This paper is organized as follows.
Section~\ref{sec:structures} provides a description of the considered stable and metastable structures followed
by an overview of the considered phase transition pathways.
The employed structures can also be found in the Supplementary informations~\cite{SupplementalMaterial}. 
The results section~\ref{sec:results} summarizes the calculated ground state energy differences of the (meta-)stable structures and their
activation energies at \SI{0}{\K} for the investigated phase transition pathways.
By calculating the Gibbs energies, pressure-temperature phase diagrams are predicted and compared to experiment. 
Furthermore we assess the temperature and pressure dependence of the activation energies.
Based on these results the experimentally observed phase transitions will be reviewed. 
Furthermore the existence of the wurtzite structure of carbon and boron nitride will be discussed.
In the course of the discussion of these results we will assess the accuracy of the various approximate electronic
structure theories.

\section{Crystal structures and phase transition pathways for carbon and boron nitride}\label{sec:structures}

\subsection{(Meta-)stable structures}

\begin{figure}[ht]
  \centering
  \subfloat[AA']{\includegraphics[width=\figw]{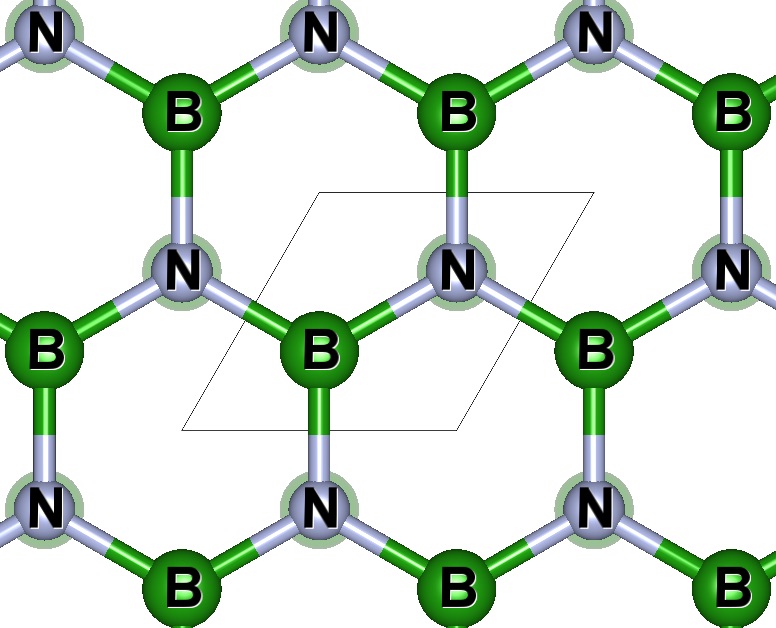}\label{fig:LAAp}}\figs
  \subfloat[AB]{\includegraphics[width=\figw]{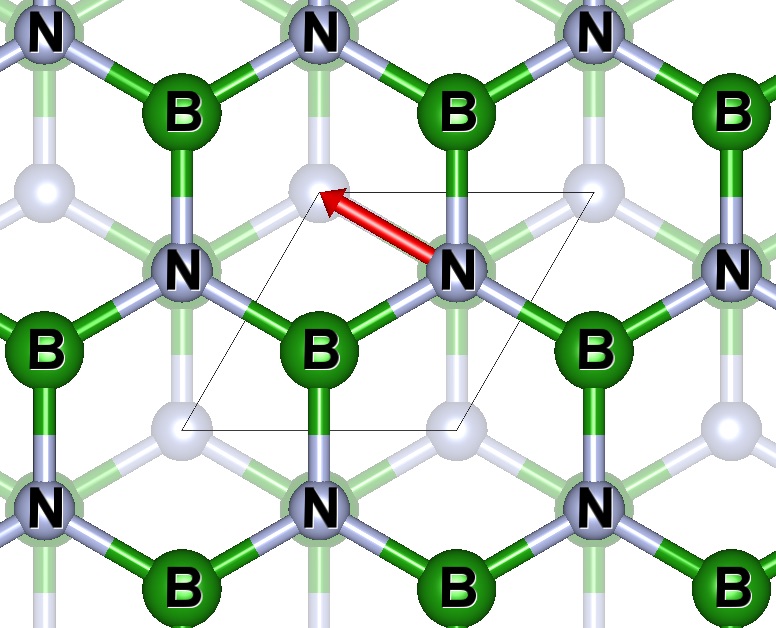}\label{fig:LAB}}\figs
  \subfloat[ABC]{\includegraphics[width=\figw]{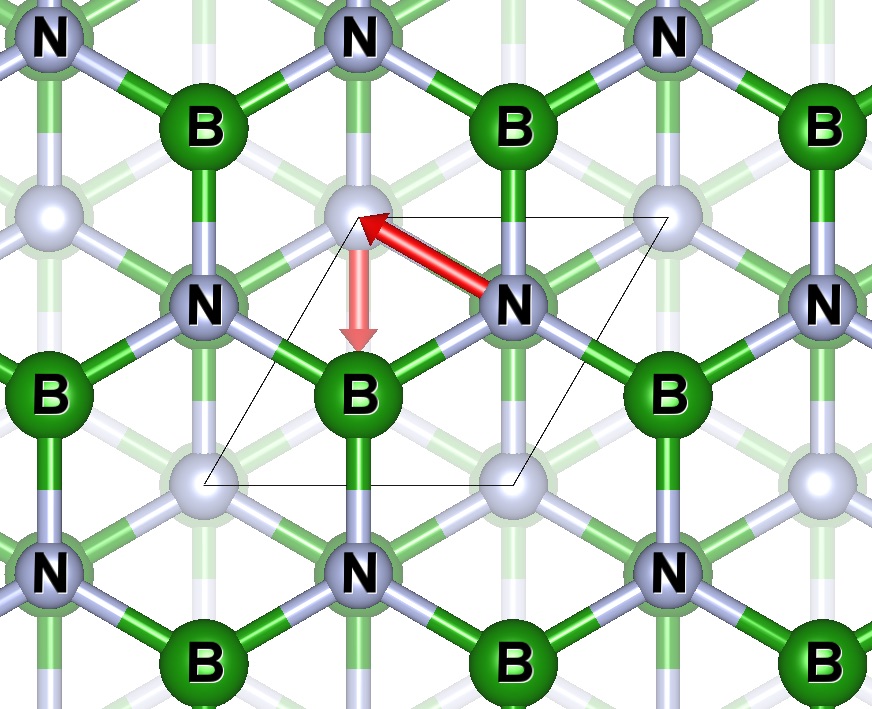}\label{fig:LABC}}\figs
  \subfloat[AD]{\includegraphics[width=\figw]{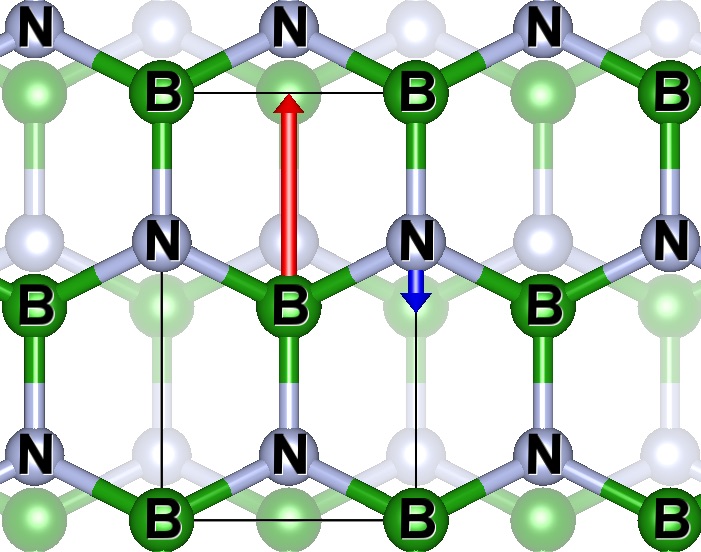}\label{fig:LAD}}\\
  \subfloat[\cBN]{\includegraphics[width=\figw]{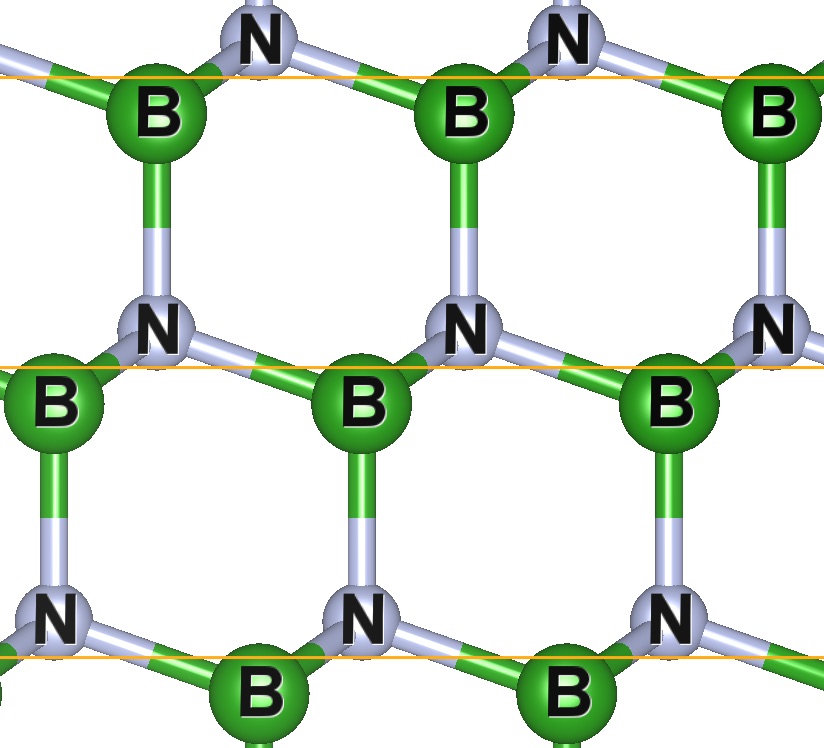}\label{fig:cBNi}}\figs
  \subfloat[\wBN]{\includegraphics[width=\figw]{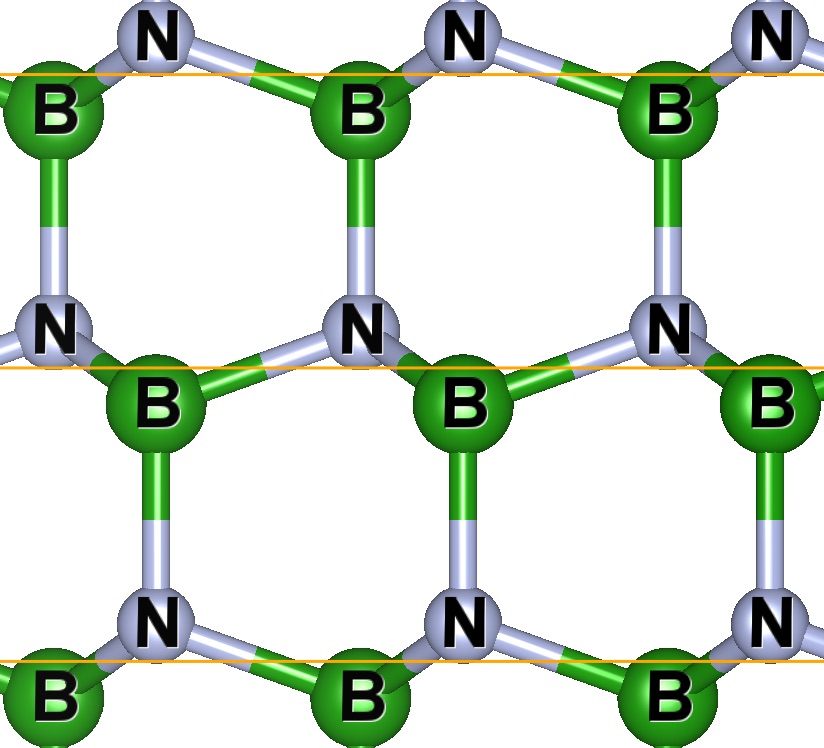}\label{fig:wBNi}}\figs
  \caption{\label{fig:layer} View along c-axis with different stacking order. A' is rotated by 60$^\circ$ compared to A.
  Translating A by the red arrows creates the B, C and D layers. Blue arrow translates the D to a B layer.
  Panel~\protect\subref{fig:LAAp} shows the  lattice vectors of the 8 atomic orthorhombic unit cell employed in
  \tref{tab:strucparam}.
  The crystal structures for carbon can easily be derived from boron nitride by substituting all B- and N-atoms by C-atoms.
  }
\end{figure}

In this work we consider the most abundant crystal structures of
carbon and boron nitride:
the graphitic and diamond-like phases.
\fref{fig:layer} illustrates the corresponding structures.
The low-density phases are graphitic with all atoms being $sp^2$ bonded and arranged in the planar honeycomb lattice with 
different stackings: AA, AB and ABC as depicted in \fref{fig:layer}.
The \GAB\ and \GABC\ 
have been observed experimentally for carbon and
can be transformed into each other by translation of the layers.
For boron nitride
the stable low-density phase is (hexagonal) \hBN. \hBN\ exhibits
an AA' stacking order, indicating the other atom types  for lattice sites on top of
each other in the direction of stacking as shown in \fref{fig:LAAp}.
We note that the AA (\GAA) stacking is unstable for carbon.

The high-density phases of carbon and boron nitride are diamond-like.
All atoms in the considered diamond-like phases can be assigned to chair or boat conformations of six-membered rings and
two different stacking orders.
For carbon and boron nitride the most stable high-pressure phases are cubic diamond (\cD) and zinc blende (\cBN),
respectively.
\cD\ and \cBN\ consist of six-membered rings of $sp^3$ bonded atoms in the chair conformation with an ABC stacking order
as shown in \fref{fig:cBNi}.
There is a second high-density phase that is referred to as wurtzite for boron nitride and hexagonal
diamond (\hD) for carbon with A\texttt{A} stacking order, exhibiting chair and boat conformations as illustrated in \fref{fig:wBNi}.
The mirror image of A is \texttt{A}, with the mirror parallel to the layer.
The chair and boat conformations are  parallel and perpendicular to the stacking, respectively. 
\hD\ is also known as lonsdaleite and serves as marker for shock impact events.

The crystal structures for C can easily be derived from BN, by substituting all B- and N-atoms with C-atoms.
This makes (rhombohedral) \rBN\ (\fref{fig:LABC}), \cBN\ (\fref{fig:cBNi}) and \wBN\ (\fref{fig:wBNi})
equivalent to \GABC, \cD\ and \hD, respectively.

Experimentally single crystals have been reported
for \hBN\ \cite{KubotaCoM2008,LuNC2015}, \cBN\ \cite{TaniguchiJoCG2001},
\GAB\ \cite{AustermanC1967} and for \cD\ \cite{KandaBJoP2000}.
There exist no single crystals for \wBN, but the XRD measurements show no sign of \cBN\ in the samples and
a small amount ($<$2\%) of the starting material \hBN\ \cite{NagakuboAPL2013}.
The synthesis of \rBN\ results in fibrous micro-crystals, but show no mixture with \hBN\ \cite{MatsuiJoMS1981}.
The \hD\ phase has been investigated in a number of theoretical as well as experimental studies in the past~\cite{FahyPRB1986,FahyPRB1987,TateyamaPRB1996,BundyC1996}.
However, recent experimental studies indicate that the previously believed samples of \hD\ are in fact \cD\
crystals that contain a large number of twins and stacking faults, creating x-ray diffraction patterns similar to the
hypothetical \hD\ \cite{NemethNC2014}.

\subsection{Structural phase transition pathways}\label{sec:structurespt}

\begin{figure}[t]
  \centering
  \subfloat[]{\includegraphics[width=0.3\columnwidth]{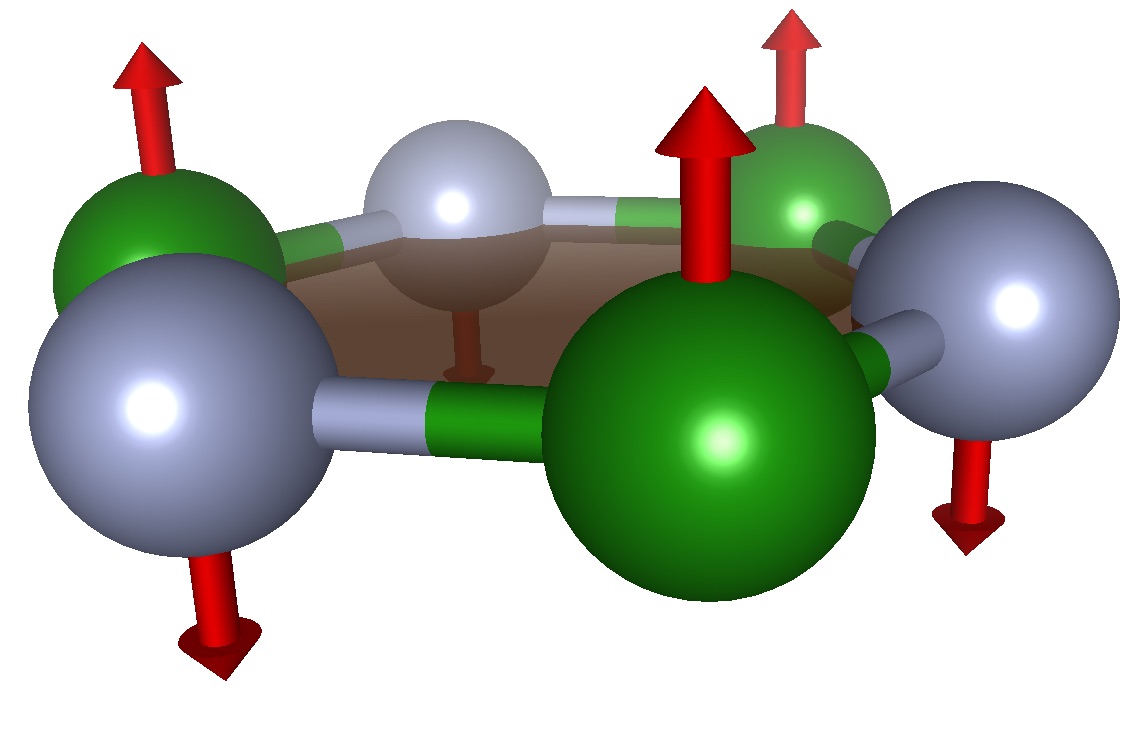}\label{fig:PG}}\figs
  \subfloat[]{\includegraphics[width=0.3\columnwidth]{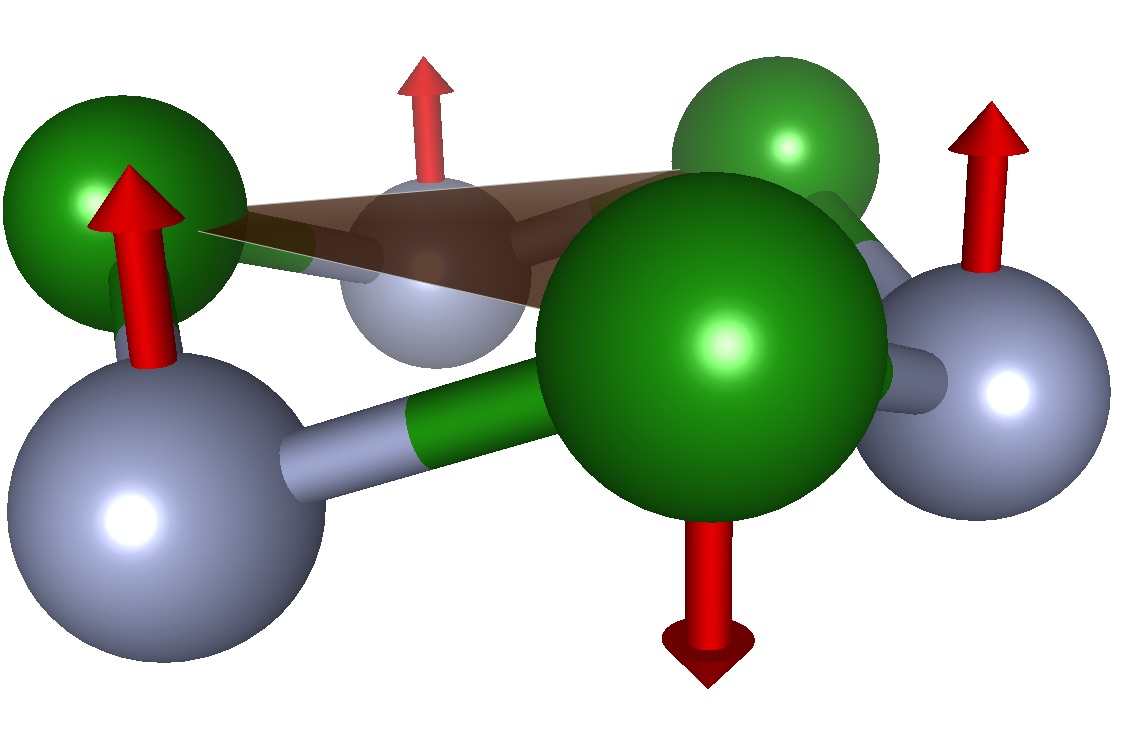}\label{fig:PD}}\figs\\
  \subfloat[]{\includegraphics[width=0.3\columnwidth]{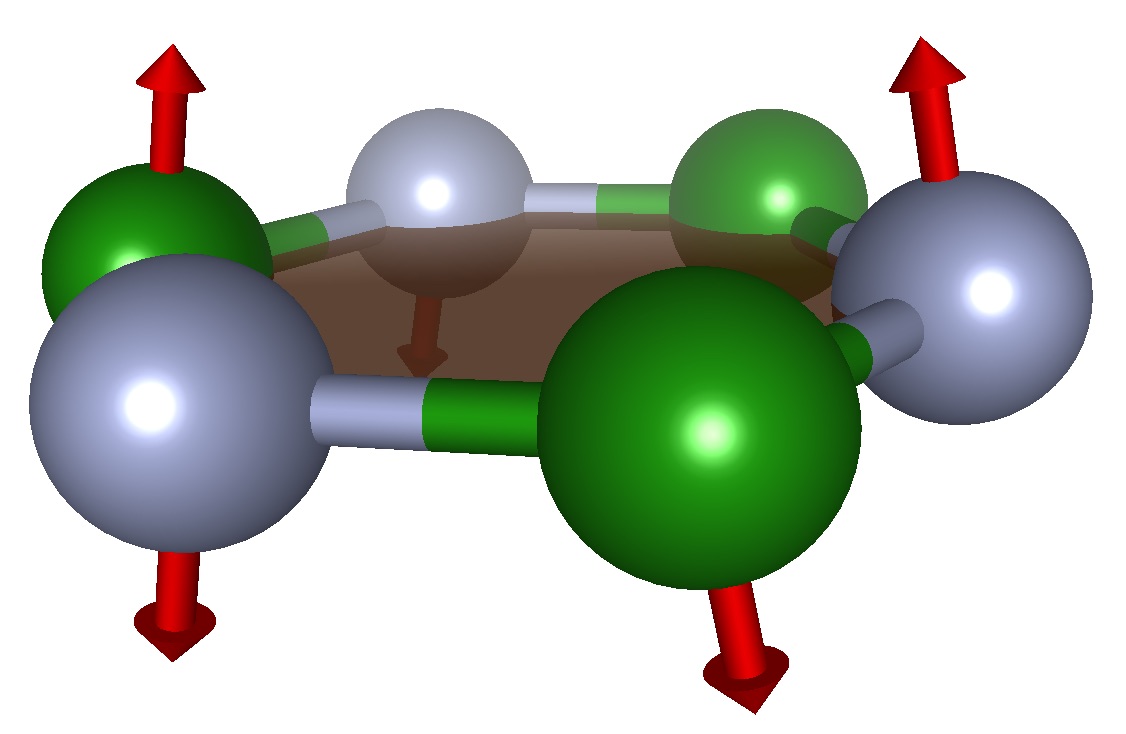}\label{fig:BG}}\figs
  \subfloat[]{\includegraphics[width=0.3\columnwidth]{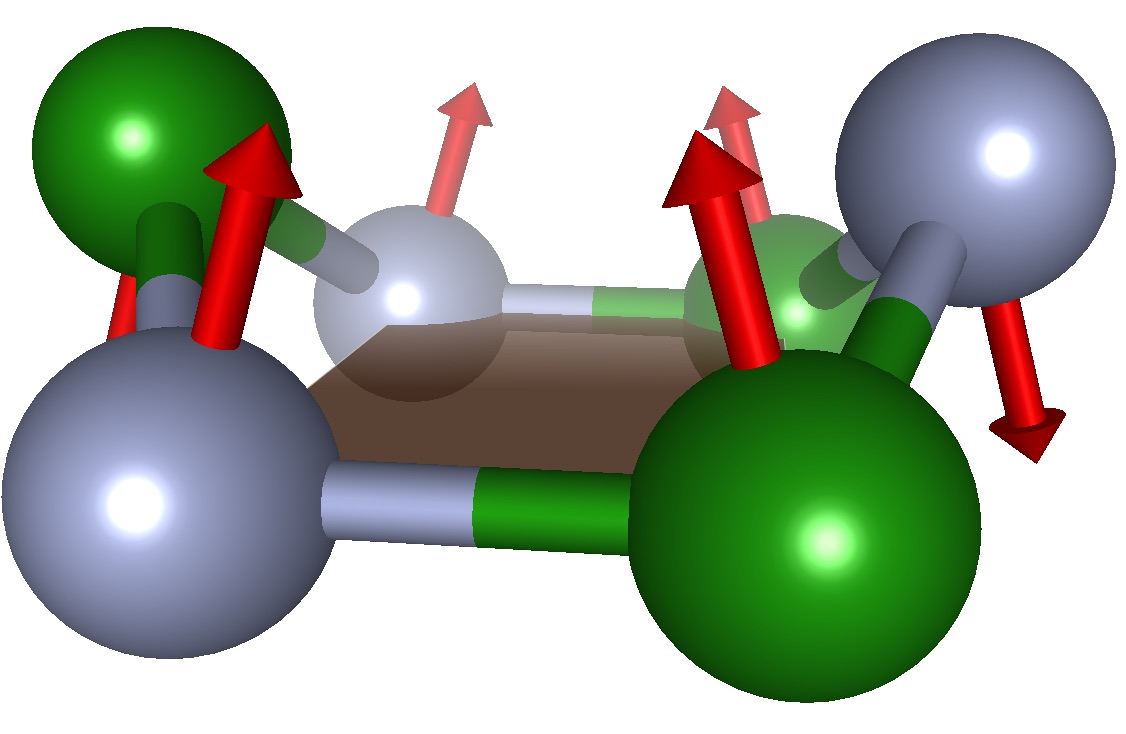}\label{fig:BD}}
  \caption{\label{fig:BPmech} Transformation from planar rings via  \protect\subref{fig:PG} \ce{->[\text{puckering}]} \protect\subref{fig:PD}
  to form the chair conformation or via \protect\subref{fig:BG} \ce{->[\text{buckling}]} \protect\subref{fig:BD} to form the boat conformation.}
\end{figure}

We now discuss the investigated structural phase transition pathways.
For the present study we keep the computational cost of the coupled cluster theory calculations low by
restricting ourselves to transition state geometries that contain at most four atoms in the unit cell.
Similar transition states have already been investigated in Refs.~\cite{FahyPRB1986,FahyPRB1987,TateyamaPRB1996,DongJoPCM2013,WentzcovitchPRB1988,WangPRL2011,YuPRB2003}.
To drive a transition from the low-density graphitic phases to the high-density diamond-like phases
the application of pressure is needed. 
Under pressure the c-axis of the graphitic phase experiences a much larger compression
than the other axes and will therefore be referred to as the compression axis. 
At high pressures the planar structure of graphite splits.
\fref{fig:BPmech} depicts two basic mechanisms by which the splitting of the planar six-membered rings present in the
honeycomb lattice occurs. The mechanisms are referred to as buckling or puckering.
Buckling and puckering creates the boat and chair conformation of six-membered rings, respectively.
We employ the following naming convention for transition states.
The first letter refers to the puckering (p) or buckling (b) mechanism and the second
letter refers to the cubic (c) or wurtzite (w)  structure corresponding to the final state
of the considered transition.

\begin{figure}[t]
  \centering
  \subfloat[\rBN]{\includegraphics[width=0.28\columnwidth]{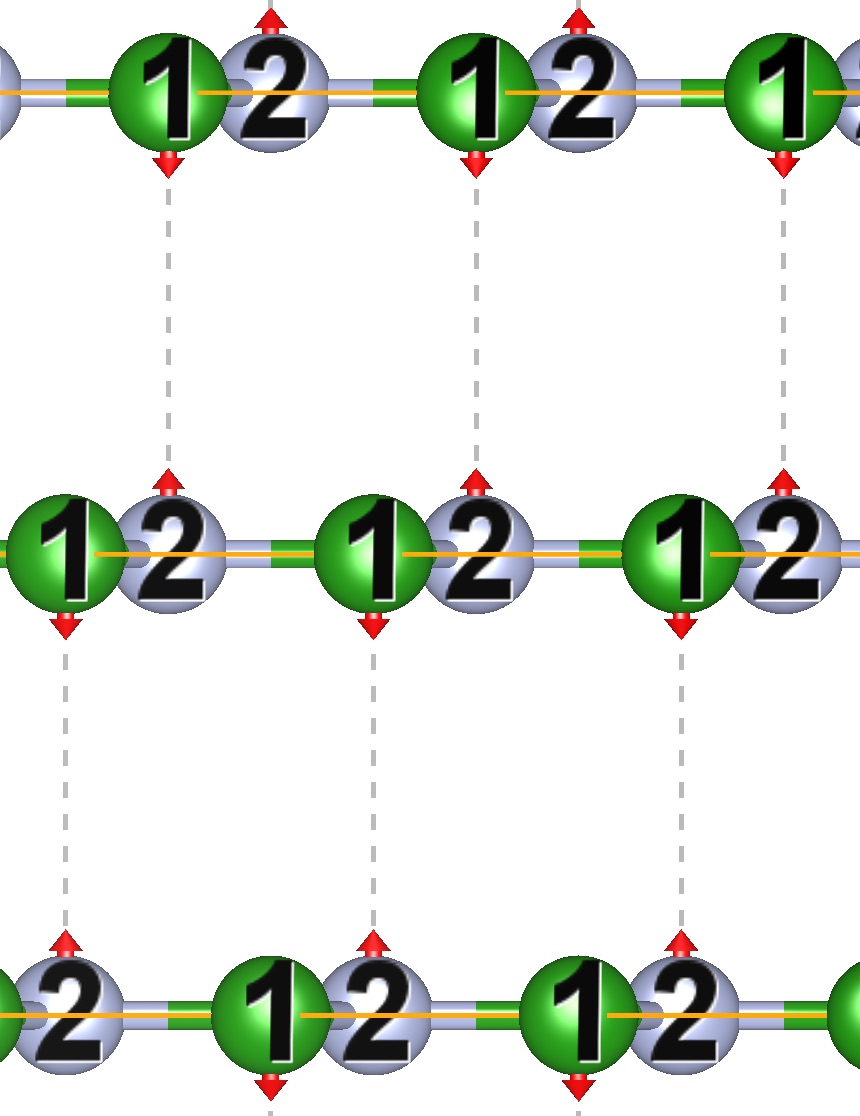}\label{fig:rBNp}}\figs
  \subfloat[\pcTS]{\includegraphics[width=0.28\columnwidth]{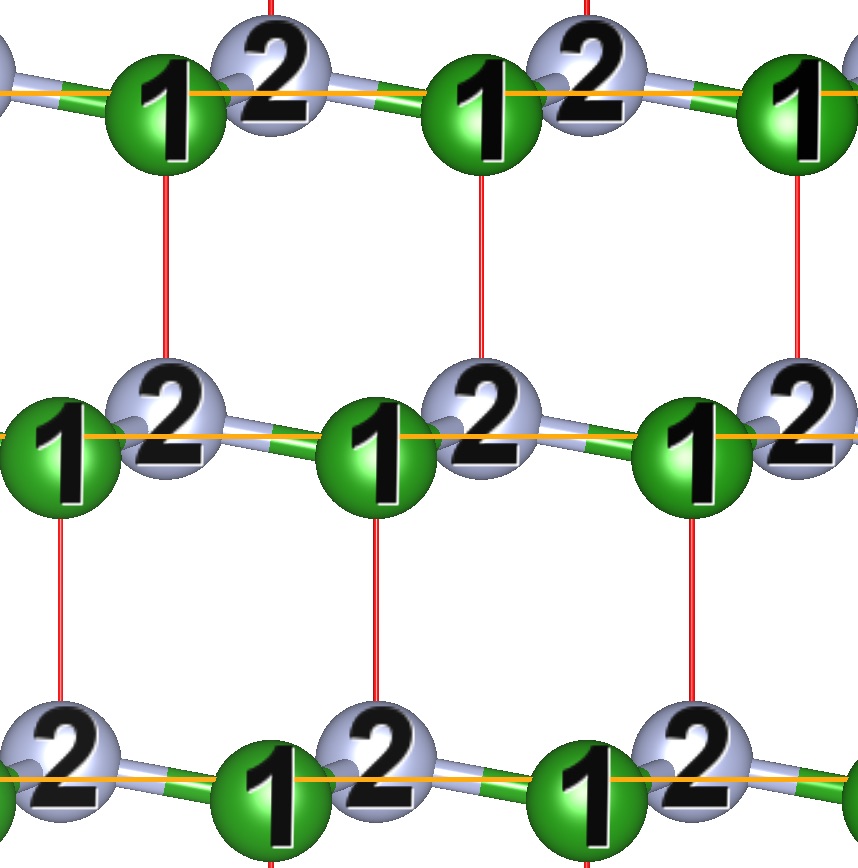}\label{fig:pcTSp}}\figs
  \subfloat[\cBN]{\includegraphics[width=0.28\columnwidth]{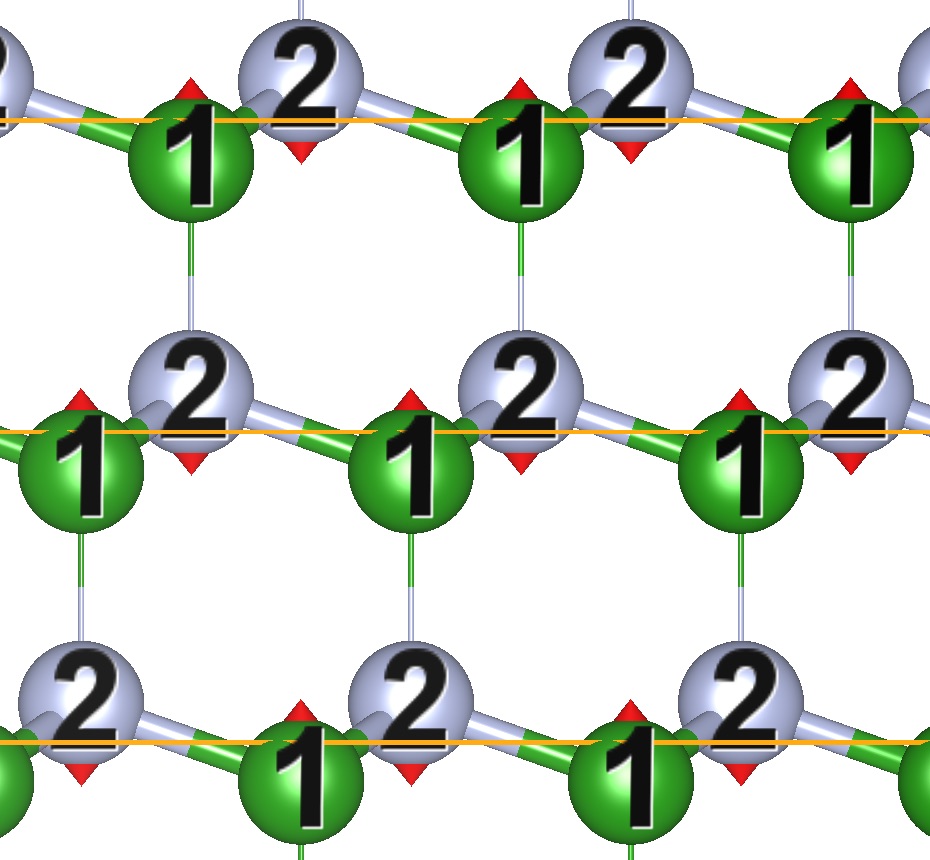}\label{fig:cBNp}}\figs
  \caption{\label{fig:pcTSpath}  \protect\subref{fig:rBNp} \ce{->[\text{\protect\subref{fig:pcTSp}}\pcTS]} \protect\subref{fig:cBNp}
  (BN: \cite{WentzcovitchPRB1988}; C: \cite{FahyPRB1986}).
  Red arrows indicate the atomic displacements and support together with the atom numbers the assignment during the phase transition.
  Dotted lines show new bonds to be formed and red lines represent strong interaction during the transition state.
}
\end{figure}

\begin{figure}[t]
  \centering
  \subfloat[\hBN]{\includegraphics[width=0.28\columnwidth]{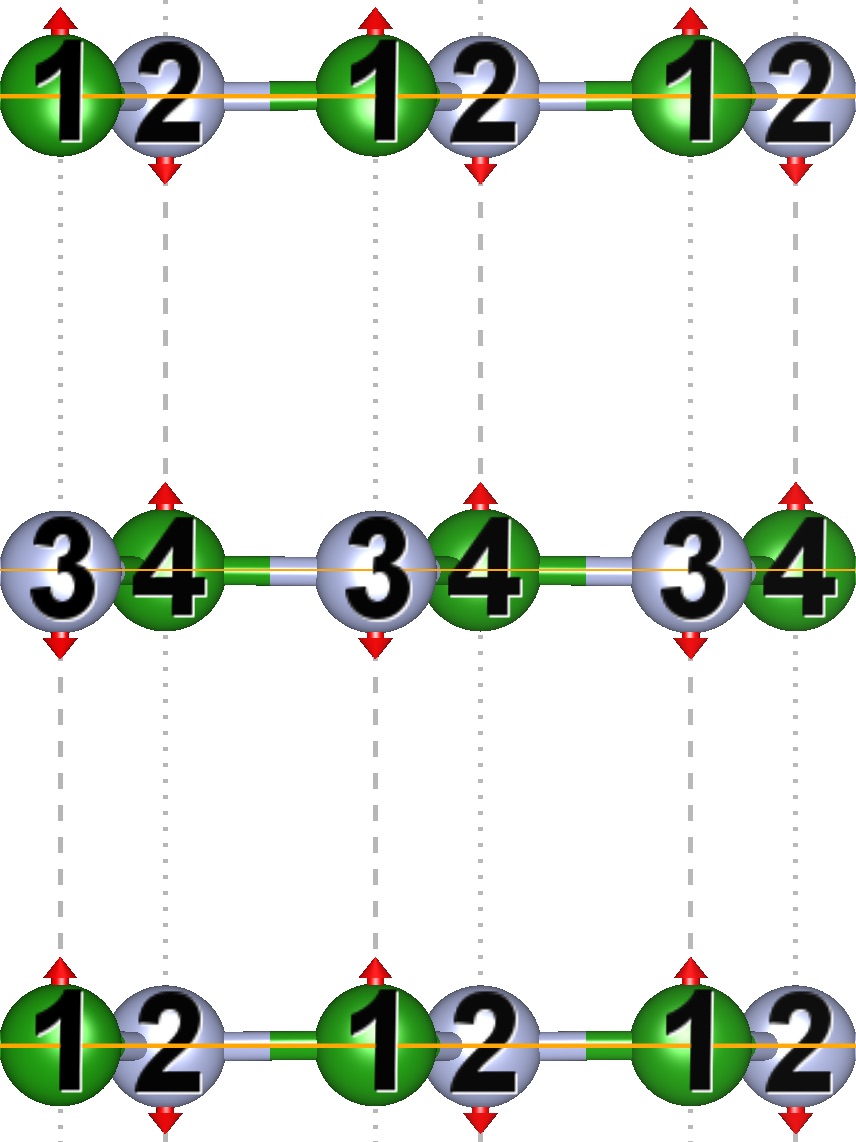}\label{fig:hBNp}}\figs
  \subfloat[\pwTS]{\includegraphics[width=0.28\columnwidth]{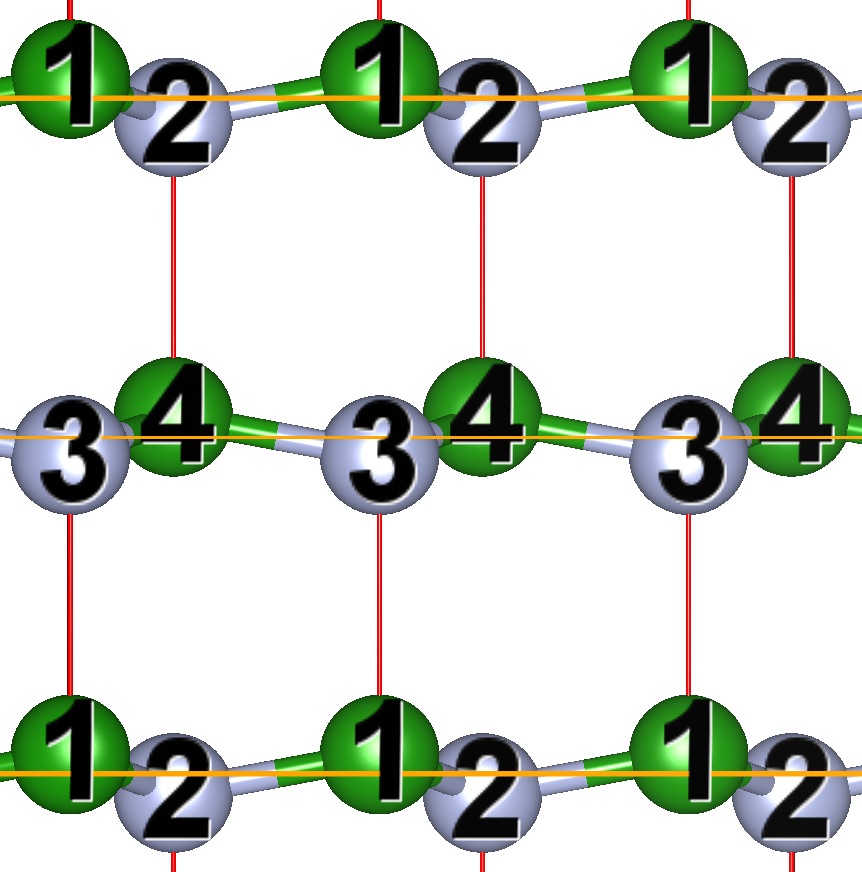}\label{fig:pwTSp}}\figs
  \subfloat[\wBN]{\includegraphics[width=0.28\columnwidth]{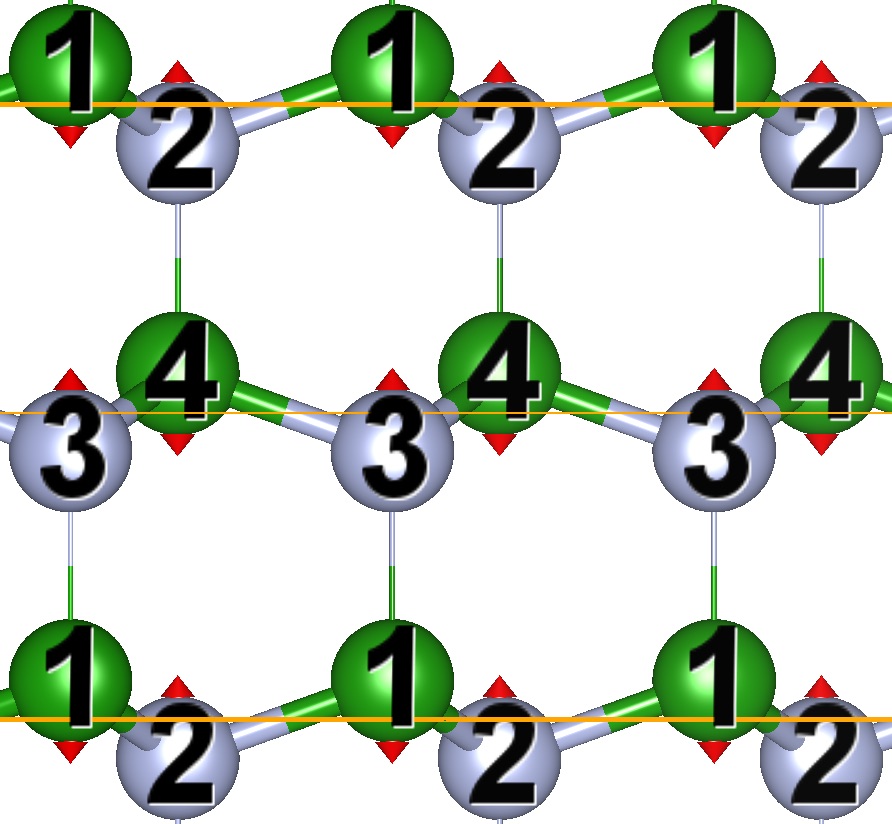}\label{fig:wBNp}}\figs
  \caption{\label{fig:pwTSpath}  \protect\subref{fig:hBNp} \ce{->[\text{\protect\subref{fig:pwTSp}}\pwTS]} \protect\subref{fig:wBNp}
  (BN: \cite{WentzcovitchPRB1988}; C: \cite{FahyPRB1987}).
  Red arrows indicate the atomic displacements and support together with the atom numbers the assignment during the phase transition.
  Dotted lines show new bonds to be formed and red lines represent strong interaction during the transition state.
}
\end{figure}
\begin{figure}[ht]
  \centering
  \subfloat[\BNAB]{\includegraphics[width=0.28\columnwidth]{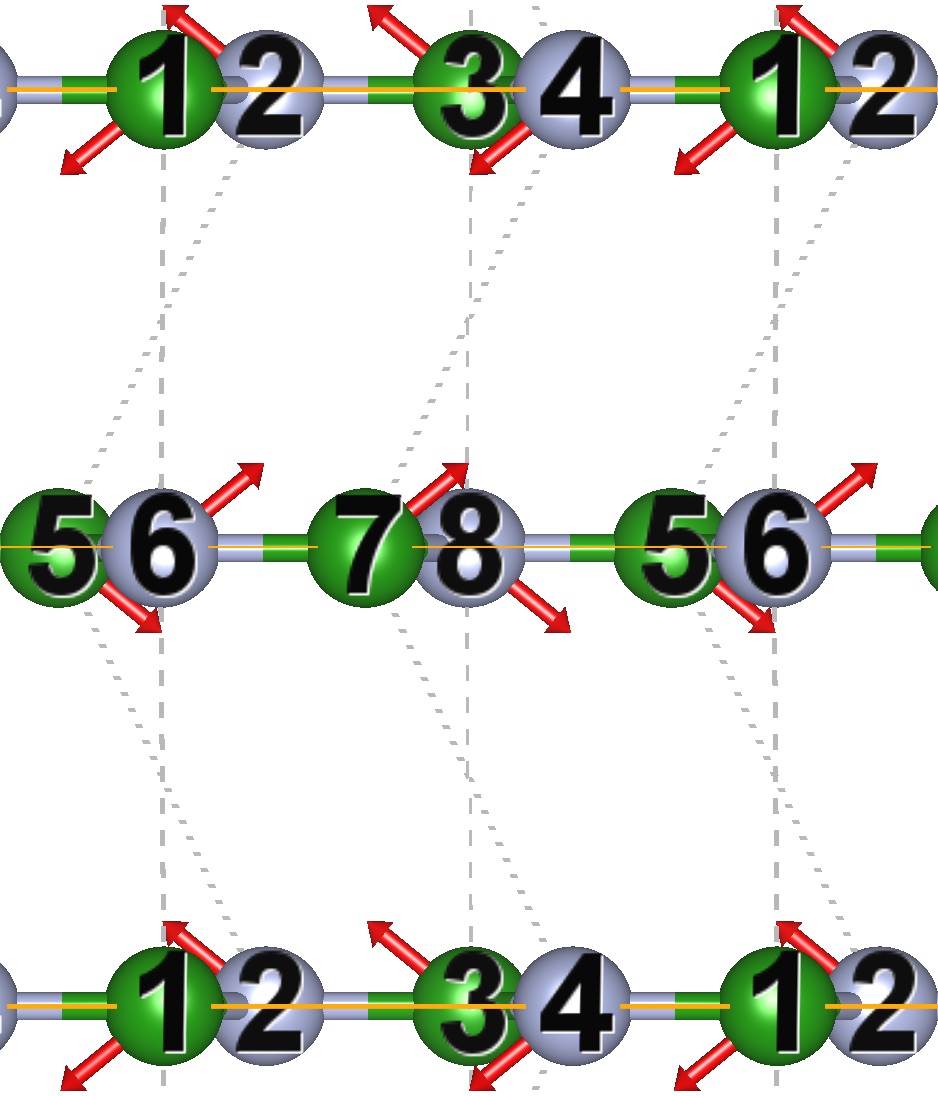}\label{fig:TGp}}\figs
  \subfloat[\bwTS]{\includegraphics[width=0.28\columnwidth]{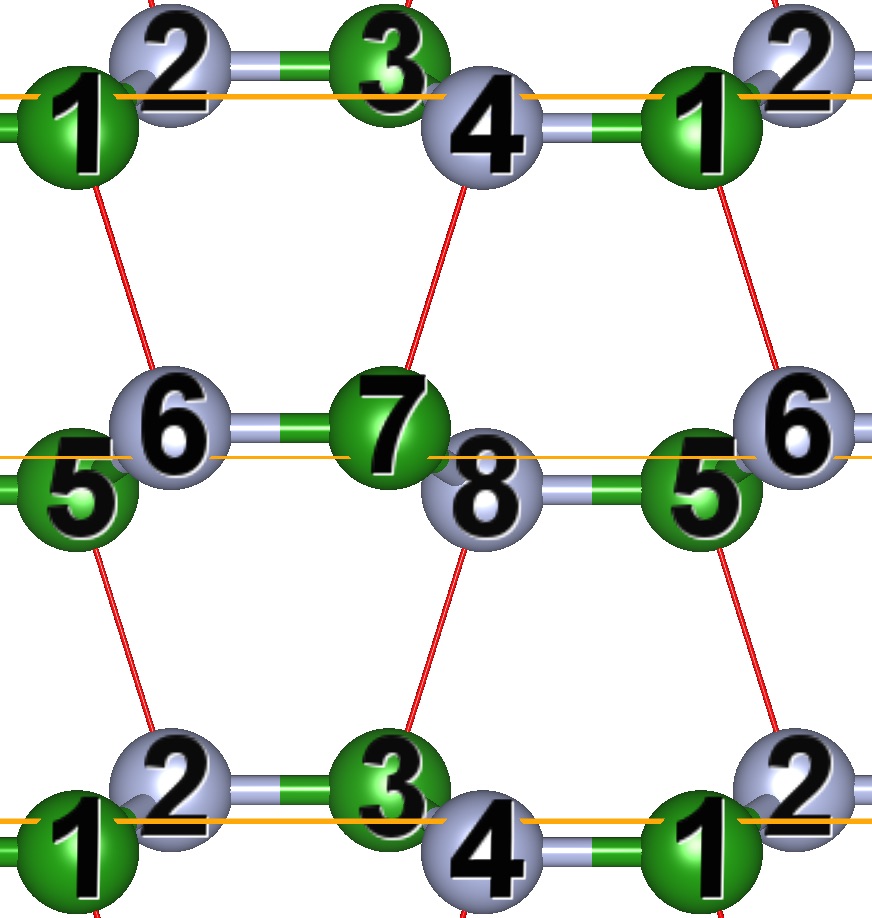}\label{fig:bwTSp}}\figs
  \subfloat[\wBN]{\includegraphics[width=0.28\columnwidth]{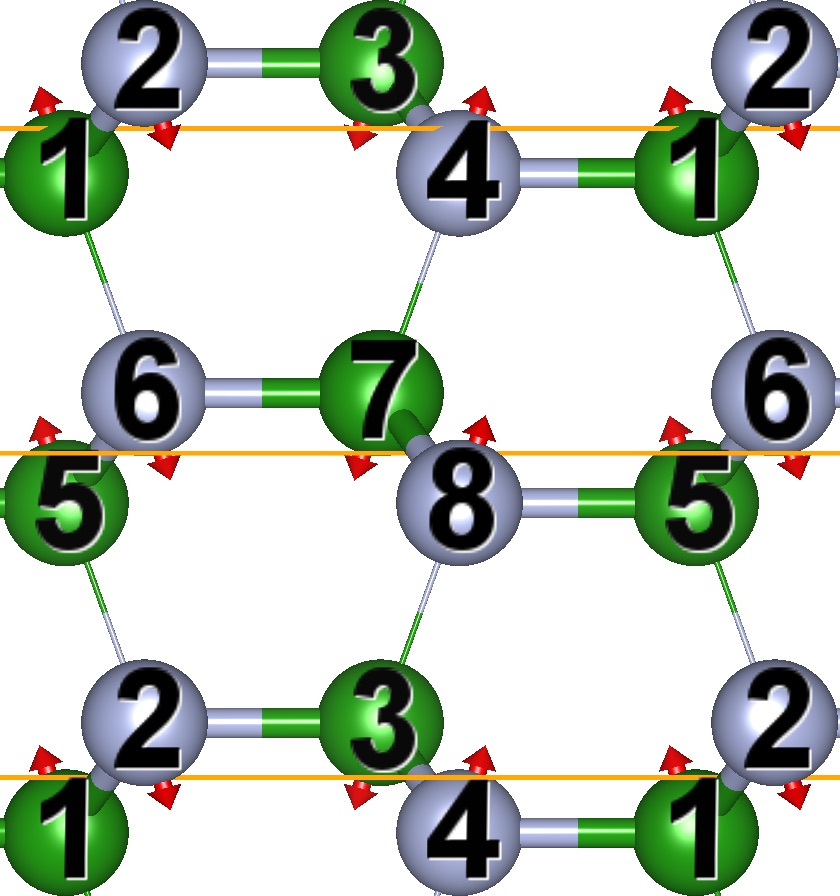}\label{fig:TDp}}\figs
  \caption{\label{fig:bwTSpath}  \protect\subref{fig:TG} \ce{->[\text{\protect\subref{fig:bwTS}}\bwTS]} \protect\subref{fig:TD}
  (BN: \cite{KurdyumovDaRM1996} for \rBN $\rightarrow$ \wBN; C: \cite{TateyamaPRB1996}).
  Red arrows indicate the atomic displacements and support together with the atom numbers the assignment during the phase transition.
  Dotted lines show new bonds to be formed and red lines represent strong interaction during the transition state.
}
\end{figure}

\pcTS\ is the transition state in the \GABC\ to \cD\ transition via the puckering mechanism.
We note that the cubic phase contains only six-membered rings in the chair conformation. Therefore
it is reasonable to consider this one transition state (\fref{fig:pcTSp}) only.

The wurtzite structure contains six-membered rings in the chair and boat conformation.
In the \pwTS\ the chair conformation is perpendicular and the boat conformation is parallel to the
compression axis (\fref{fig:pwTSp}), whereas in the \bwTS\ the orientations are switched (\fref{fig:bwTSp}).
By comparing the resulting structures from the corresponding transition pathways one can see that
the c-axis of \wBN\ is rotated by 90$^\circ$ (\fref{fig:wBNp} and \ref{fig:TDp}). 

\begin{figure}[t]
  \centering
  \subfloat[\wBN]{\includegraphics[width=0.28\columnwidth]{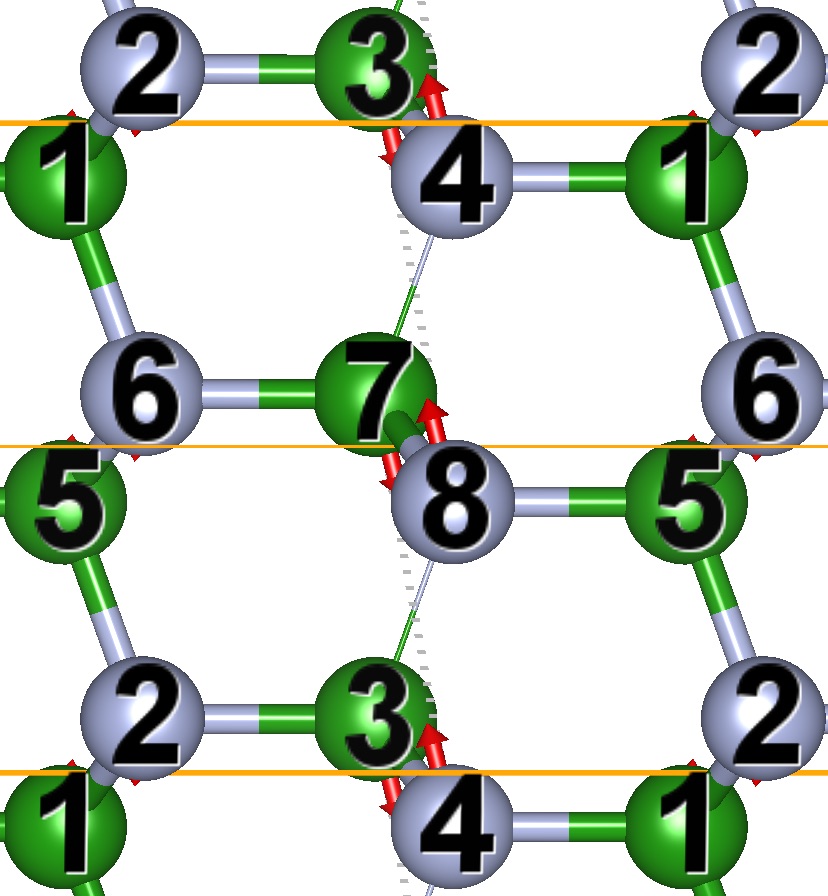}\label{fig:wBNcp}}\figs
  \subfloat[\lpcTS]{\includegraphics[width=0.28\columnwidth]{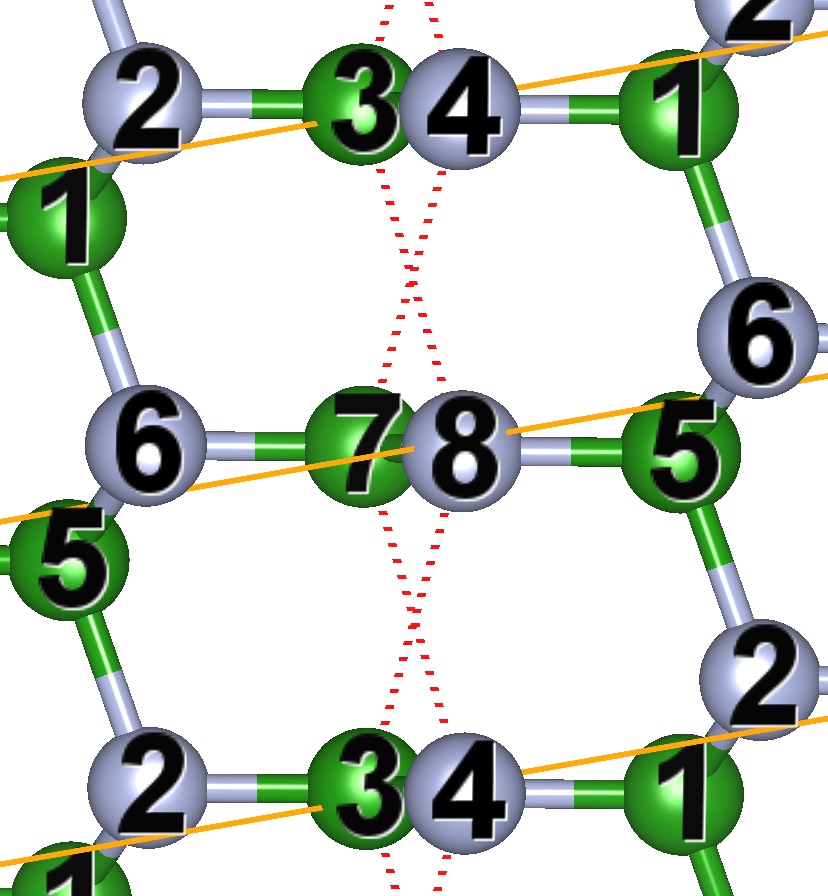}\label{fig:lpcTSp}}\figs
  \subfloat[\cBN]{\includegraphics[width=0.28\columnwidth]{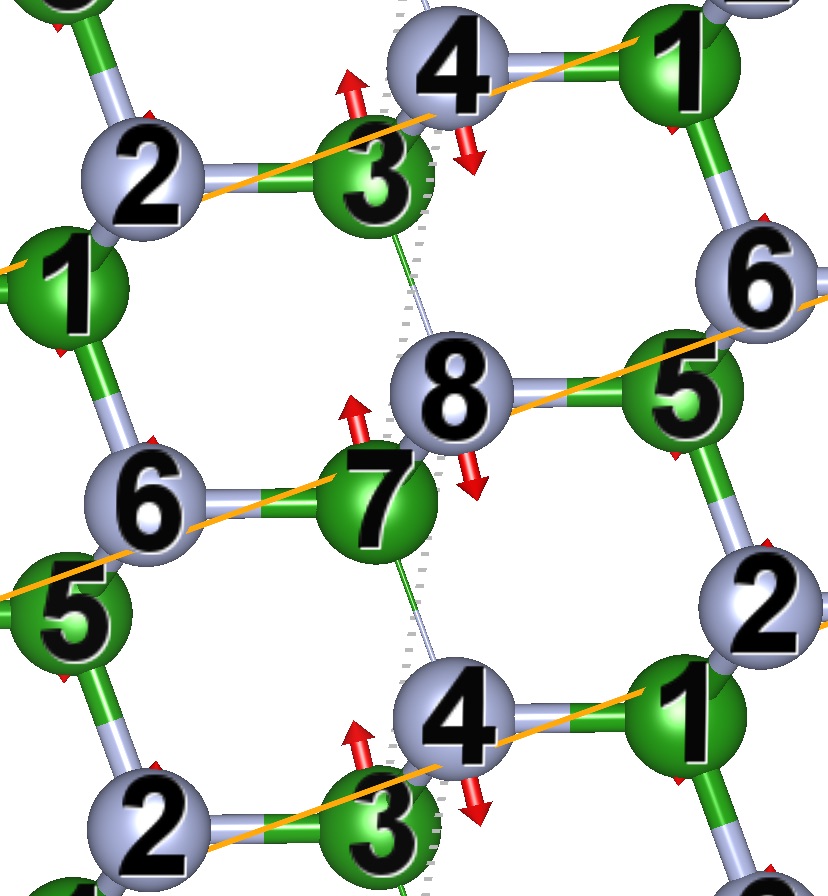}\label{fig:cBNcp}}\figs
  \caption{\label{fig:lpcTSpath}
  Red arrows indicate the atomic displacements and support together with the atom numbers the assignment during the phase transition.
  Dotted lines show new bonds to be formed and red lines represent strong interaction during the transition state.
}
\end{figure}

We also consider the transition from  \wBN\ to \cBN,
which occurs in a stepwise layer-to-layer rearrangement
through 4H intermediate structures and will be referred to as \lpcTS\ in this work \cite{BritunJoMS1993}.
During this transformation the boat conformation along the c-axis of the \wBN\ structure transforms into the chair
conformation, while the six-membered rings perpendicular to the c-axis break apart (e.g. atoms 3+8 bottom layers) and rebond differently
(e.g. atom 4+7 bottom layers) (\fref{fig:wBNcp} \ce{->[\text{\ref{fig:lpcTSp}}]} \ref{fig:cBNcp} with plane (001)$\lt{w}
\parallel$ (111)$\lt{c}$ and direction [10$\overline{1}$0]$\lt{w} \parallel$ [112]$\lt{c}$). 

In total we consider four different transition states including
\pcTS, \pwTS, \bwTS\ and \lpcTS\ for carbon and boron nitride allotropes.
In the case of carbon all lattice sites are occupied by the same atomic species.

The geometries of the transition states have been determined as follows.
For the four atomic unit cells of the \pcTS\ and \pwTS\ three degrees of freedom were considered:
the intra and inter layer bond distance and the angle in between.
The first-order saddle point on the corresponding potential energy surface defines
the transition state geometry as well as its energy. Due to the small number of considered
degrees of freedom a sweeping algorithm was sufficient to determine the \pcTS\ and \pwTS.
Determining the  \bwTS\ and \lpcTS\ is slightly more complicated due to the larger number of degrees of freedom.
For \bwTS\ the $z$-coordinate of the interlayer bond distance $R$, the horizontal lattice vectors in \fref{fig:bwTSp},
the out-of-plane displacement of the atoms and the in-plane coordinates of all atoms were considered as
degrees of freedom.
The \bwTS\ was obtained applying a sweeping algorithm to all degrees of freedom except for the
in-plane coordinates of all atoms and the lattice vectors that were optimized by relaxing the
structures accordingly for a given unit cell.
The  \lpcTS\ was obtained in a similar manner.
A sweeping algorithm was employed to calculate energies for all coordinates and cell parameters interpolated
linearly between the initial (\wBN) and the final (\cBN) structure. 
For each interpolated structure all cell parameters and atomic positions were allowed to relax
while keeping only the vertical-coordinate of the atoms in \fref{fig:wBNcp} frozen.

\subsection{Crystal lattice parameters}

\begin{table*}[t]
  \begin{center}
    \caption{\label{tab:strucparam} Structural parameters of the eight atomic monoclinic type unit cell ($\beta$~=~$\gamma$~=~\SI{90}{\degree}) including
    interlayer distance $d$, the forming bond distance $R$ marked red in the transition state of \fref{fig:pcTSpath}~--~\ref{fig:lpcTSpath} and its
    corresponding bond length in the other structures, atomic volume $V$ and density $\rho$.   
    See text for further details. }
    \begin{tabular}{r k{3.3}k{3.3}k{3.3} k{3.3}k{3.3}k{3.3}k{3.3} k{3.3}k{3.3}}
      \hline
      \hline
      Carbon & \multicolumn{1}{c}{\GABC} & \multicolumn{1}{c}{\GAB} & \multicolumn{1}{c}{\GAA} & \multicolumn{1}{c}{\pcTS} & \multicolumn{1}{c}{\pwTS} & \multicolumn{1}{c}{\bwTS} & \multicolumn{1}{c}{\lpcTS} & \multicolumn{1}{c}{\cD} & \multicolumn{1}{c}{\hD}\\
      \hline
      $|\vec{a}|$ [\AA] & 2.446 & 2.446 & 2.445 & 2.465 & 2.461 & 2.471 & 2.456 & 2.498 & 2.484\\
      $|\vec{b}|$ [\AA] & 4.236 & 4.236 & 4.236 & 4.270 & 4.263 & 4.164 & 4.229 & 4.326 & 4.302\\
      $|\vec{c}|$ [\AA] & 6.735 & 6.590 & 7.171 & 4.899 & 4.780 & 4.835 & 4.705 & 4.326 & 4.137\\
      $\alpha$ & 77.9 & 90.0 & 90.0 & 73.1 & 90.0 & 90.0 & 80.3 & 70.5 & 90.0\\
      d  [\AA] & 3.292 & 3.295 & 3.585 & 2.344 & 2.390 & 2.418 & 2.319 & 2.039 & 2.069\\
      R  [\AA] & 3.292 & 3.295 & 3.585 & 2.058 & 2.108 & 2.100 & 2.472 & 1.530 & 1.549\\
      V  [\si{\cubic\angstrom\per\atom}] & 8.528 & 8.535 & 9.285 & 6.167 & 6.268 & 6.220 & 6.021 & 5.510 & 5.526\\
      $\rho$ [\si{\gram\per\cubic\centi\metre}] & 2.417 & 2.415 & 2.220 & 3.343 & 3.289 & 3.315 & 3.424 & 3.742 & 3.731\\
      \hline
      Boron nitride & \multicolumn{1}{c}{\rBN} & \multicolumn{1}{c}{\BNAB} & \multicolumn{1}{c}{\hBN} & \multicolumn{1}{c}{\pcTS} & \multicolumn{1}{c}{\pwTS} & \multicolumn{1}{c}{\bwTS} & \multicolumn{1}{c}{\lpcTS} & \multicolumn{1}{c}{\cBN} & \multicolumn{1}{c}{\wBN}\\
      $|\vec{a}|$ [\AA] & 2.488 & 2.488 & 2.488 & 2.506 & 2.506 & 2.511 & 2.513 & 2.532 & 2.524\\
      $|\vec{b}|$ [\AA] & 4.309 & 4.309 & 4.310 & 4.341 & 4.341 & 4.219 & 4.282 & 4.386 & 4.371\\
      $|\vec{c}|$ [\AA] & 7.069 & 6.458 & 6.491 & 4.932 & 4.742 & 4.900 & 4.707 & 4.386 & 4.176\\
      $\alpha$ & 77.5 & 90 & 90 & 72.9 & 90 & 90 & 79.5 & 70.5 & 90\\
      d  [\AA] & 3.229 & 3.229 & 3.246 & 2.357 & 2.371 & 2.450 & 2.315 & 2.068 & 2.088\\
      R  [\AA] & 3.229 & 3.229 & 3.246 & 2.088 & 2.102 & 2.128 & 2.447 & 1.551 & 1.564\\
      V  [\AA$^3$] & 9.246 & 8.653 & 8.700 & 6.411 & 6.448 & 6.489 & 6.227 & 5.742 & 5.759\\
      $\rho$ [\si{\gram\per\cubic\centi\metre}] & 2.230 & 2.382 & 2.370 & 3.216 & 3.197 & 3.177 & 3.311 & 3.590 & 3.580\\
      \hline
      \hline
    \end{tabular}
  \end{center}
\end{table*}

\begin{table}[t]
  \begin{center}
    \caption{\label{tab:expdiff} Relative difference to experimental
    values \cite{LynchTJoCP1966,YoshiasaJJoAP2003,SlackJoAP1975,FurthmuellerPRB1994,NagakuboAPL2013}
    of lattice vectors $|\vec{a}|$ and $|\vec{c}|$ for stable structures of carbon an BN. In the case of \wBN, \hBN, \hD\ and \GAB\ we compare
    to a range of experimental values.}
    \begin{tabular}{rrrrrrr}
      \hline \hline
      & \multicolumn{1}{c}{\cBN} & \multicolumn{1}{c}{\wBN} & \multicolumn{1}{c}{\hBN} & \multicolumn{1}{c}{\cD} & \multicolumn{1}{c}{\hD} & \multicolumn{1}{c}{\GAB} \\
      \hline
      $|\vec{a}|$ [\%] & 0.94 & \numrange{1.03}{1.04} & \numrange{0.47}{0.63} & 0.97 & \numrange{1.04}{0.96} & \numrange{0.58}{0.62}\\
      $|\vec{c}|$ [\%] &  & \numrange{0.63}{1.21} & \numrange{2.53}{2.55} &   & \numrange{-0.62}{1.10} & \numrange{0.78}{1.75} \\
      \hline
      \hline
    \end{tabular}
  \end{center}
\end{table}

\tref{tab:strucparam} summarizes the lattice parameters of the employed geometries for the
low- and high-density phases and transition states for carbon and boron nitride allotropes.
These parameters have been optimized using DFT in the LDA.
This is necessary because forces are not
yet implemented in the employed coupled cluster theory code. We believe that the LDA
provides sufficiently accurate structures compared to experiment that allow for an unbiased comparison
between the employed electronic structure theories and to experiment.
We note that the LDA lattice parameters deviate by about \SI{1}{\percent} only from experiment
even for the lattice vector $|\vec{c}|$ parallel to the compression axis as summarized in \tref{tab:expdiff}.
The only exception is \hBN, where the deviation is slightly larger.
To allow for a direct comparison between the lattice parameters of the (meta-)stable structures as well as
transition states we consider an eight atomic unit cell with monoclinic lattice vectors $\vec{a}$, $\vec{b}$ and $\vec{c}$ such that
$|\vec{a}|$~$\neq$~$|\vec{b}|$~$\neq$~$|\vec{c}|$, $\beta$~=~$\gamma$~=~\SI{90}{\degree} and $\alpha$ can also be \SI{90}{\degree}.
The $\vec{a}\vec{b}$ plane can be seen in \fref{fig:LAD}. 
The length of $\vec{a}$ corresponds to the width of a honeycomb ring and the vector points out of plane in
\fref{fig:pcTSpath} -- \ref{fig:lpcTSpath}.
$\vec{b}$ points from left to right and $\vec{c}$ from bottom to top and spans across two layers.
The ratio $|\vec{a}|$:$|\vec{b}|$ is 1:$\sqrt{3}$ for all structures except for \bwTS\ and \lpcTS.
In these cases the ratio is larger by up to 3~\%.
We point out that in \lpcTS\ $R$ is much larger compared to the other transition states.
However, $R$ refers to all bonds between the layers in the other transition states, whereas this not the case for \lpcTS. 
In \lpcTS\ additional interlayer bonds exist with a bond length of \SI{1.654}{\angstrom} (1+6 or 2+5 in \fref{fig:lpcTSp}).
For carbon (BN) the average of these two bond lengths is \SI{2.063}{\angstrom} (\SI{2.050}{\angstrom}),
respectively and comparable to $R$ of the other transitions states. The employed structures
can be found in the Supplementary informations~\cite{SupplementalMaterial}.

\section{Methods} \label{sec:method}
\subsection{Density functional and Hartree--Fock theory}

All electronic structure calculations have been performed using the projector augmented wave (PAW)
method~\cite{BlochlPRB1994} as implemented in the Vienna $ab-initio$ simulation package (\textsc{VASP})
\cite{KresseJoPCM1994,KressePRB1996}. 
We present results obtained using some of
the most widely-used $ab-initio$ methods to approximate the exchange and correlation energy in the framework
of DFT.
These methods include the LDA functional as parametrized by Perdew-Zunger~\cite{PhysRevLett.45.566,PhysRevB.23.5048}, the GGA functional
as parametrized by Perdew-Burke-Ernzerhof (PBE)~\cite{PhysRevLett.77.3865}, the dispersion corrected PBE functional
using the many-body dispersion energy method (PBE+MBD)~\cite{PhysRevLett.108.236402,Bucko2016},
the meta-GGA as parametrized for the SCAN functional~\cite{PhysRevLett.115.036402} and the hybrid density functionals
PBE0~\cite{doi:10.1063/1.472933}, Becke-3-parameter-Lee-Yang-Parr (B3LYP)~\cite{doi:10.1063/1.464913} and
Heyd-Scuseria-Ernzerhof (HSE06)~\cite{doi:10.1063/1.1564060,doi:10.1063/1.2404663}.
We note that we have chosen only a small selection of functionals that could be considered.

The B 2$s^22p^1$, N 2$s^22p^3$ and C 2$s^22p^2$ states have been treated as valence states in all calculations.

The geometries have been relaxed until the forces on all atoms are smaller than \SI{e-5}{\eV\per\angstrom}.
The total energies have been converged using the self-consistent field approach to within \SI{e-8}{\eV}.
For all DFT and HF calculations we employed a 16 atom supercell and a $4\times4\times4$ Monkhorst-Pack
$k$-point mesh.
The corresponding supercell structures are summarized in the supplementary informations.
The kinetic energy cutoff for the plane wave basis set was set to \SI{1000}{\eV}.
We note that smaller kinetic energy cutoffs would have sufficed but these calculations do not consitute
a computational bottle neck compared to the more expensive coupled cluster theory calculations.

The phonon calculations have been performed using the \textsc{Phonopy} code~\cite{TogoSM2015},
creating the displacements within a $2\times2\times2$ supercell of the 16 atom cell.
The forces are calculated using \textsc{VASP} and the LDA.
These calculations employed a kinetic energy cutoff of \SI{800}{\eV} and a $2\times2\times2$ $k$-mesh.

\subsection{Quantum chemical wavefunction theories}

Results obtained using post-Hartree--Fock methods have been converged with respect
to several computational parameters including energy cutoffs defining the plane wave basis sets,
the number of virtual orbitals and the $k$-mesh.
We have employed kinetic energy cutoffs of 500~eV for definining the orbital plane wave basis set and 
300~eV for definining an auxiliary basis set that is used in the calculation of electron repulsion integrals required
in post-HF methods.
For the twist averaging technique we have employed a 4$\times$4$\times$4 $k$-mesh.
Furthermore 14 unoccupied orbitals have been used
per atom.
These parameters ensure a convergence of the energy difference between graphite and diamond to within
a few meV per atom.
The same parameters have been employed in Ref.~\cite{fs3d}. 
For the virtual orbital space we employ MP2 natural orbitals that are obtained using a procedure
outlined in Ref.~\cite{GrueneisJoCTaC2011}.
Our estimates of the remaining basis set incompleteness error indicate that the
energy difference between carbon diamond and graphite should be converged to
within approximately \SI{4}{\meV\per\atom} as shown in Fig.2 of Ref.~\cite{fs3d}. A similar level of accuracy is expected for the other energy differences.

\subsubsection{Finite size errors}\label{sec:fseffects}

We stress that the convergence of ground state energies obtained using post-HF methods such as MP2 theory with respect to
the employed $k$-mesh or supercell size is slower than for their DFT counterparts.
Wavefunction based methods account for non-local electronic correlation effects explicitly
and therefore the observed interatomic interactions such as van der Waals forces lead to a slower
rate of convergence with respect to the employed $k$-mesh.
Recently we have introduced a finite size correction scheme that allows for accelerating the rate of convergence
for periodic systems.
We refer to results obtained using the finite size correction scheme by employing the following naming convention.
Corrected MP2 and CCSD results are referred to as MP2-TA-FS and CCSD-TA-FS, respectively.
TA and FS stand for twist averaging and an interpolation method, respectively.
For the perturbative triples (T) correction on top of CCSD-TA-FS, we employ the twist averaging technique only.
As such CCSD(T)-TA refers to CCSD-TA-FS plus the (T)-TA contribution.
The improved $k$-mesh and supercell size convergence of CCSD-TA-FS was demonstrated and discussed in Ref.~\cite{fs3d}.
If not stated otherwise, all MP2, CCSD and CCSD(T) results in this work include the finite size corrections.

In the present work we employ box plots to depict finite size errors
in Figs.~\ref{fig:DGmethod}, \ref{fig:CTSGmethod}, \ref{fig:CTSpcmethod}
\ref{fig:CRmethod}, \ref{fig:BNTSGmethod} and \ref{fig:BNTSpcmethod}.
The box plots show the distribution of the obtained results for a set of different $k$-meshes.
We stress that finite size errors must not be confused with stochastic errors.
However, small error bars indicate that results are not affected significantly by the size of the
employed $k$-mesh and can therefore be considered converged with respect to the $k$-mesh density.
The employed box plots mark the largest and lowest value, the second and third quartile as well as the mean value.
The bars in Figs.~\ref{fig:DGmethod}, \ref{fig:CTSGmethod}, \ref{fig:CTSpcmethod}
\ref{fig:CRmethod}, \ref{fig:BNTSGmethod} and \ref{fig:BNTSpcmethod}
mark the value with the largest $k$-point mesh.
Some results for certain $k$-meshes differ significantly from the other $k$-meshes and were marked as outlier
with a ''+``. We find that very anisotropic Brillouin zone sampling using $k$-meshes with one $k$-point
along one direction only yields results that are considered outliers. 

For the results depicted in Figs.~\ref{fig:DGmethod} and~\ref{fig:CRmethod} we employ two-atomic unit cells and the following
$k$-meshes:  $2\times2\times2$, $2\times2\times4$, $3\times3\times2$ and $3\times3\times3$.

For the calculations of the barrier heights shown in Figs.~\ref{fig:CTSGmethod},~\ref{fig:CTSpcmethod} and
Figs.~\ref{fig:BNTSGmethod},~\ref{fig:BNTSpcmethod} we have employed four and eight atomic unit cells.
The following $k$-meshes were employed to sample the Brillouin zone of the four atomic unit cells
\GABC/\hBN, \pcTS, \pwTS\ (\bwTS\ and \lpcTS\ in parentheses):
$2\times2\times2$ ($1\times2\times4$), $3\times3\times1$ ($1\times3\times3$)
and $3\times3\times2$ ($2\times3\times3$).
We note that two transition states (\bwTS\ and \lpcTS) need four atoms per layer to be described correctly and have only one layer. 
Therefore the $c$-axis is just half as long and the $a$-axis is doubled compared to the other cells. Consequently the
$k$-mesh was adjusted appropriately.
For the eight atomic unit cell we have employed a $1\times3\times2$ $k$-mesh.

\subsection{Thermodynamic properties}

For the calculation of pressure-temperature phase diagrams we need to compute the Gibbs energies ($G$) of all
phases.
$G$ is defined as the sum of the ground state energy, as obtained from DFT or a similar approach ($E$\lt{tot}),
all entropic contributions and the $pV$ term.
The vibrational ($F$\lt{vib}) contribution is the largest entropy related contribution.
Using the finite displacement method a phonon density of state ($D(\omega)$) has been obtained for the
frequency range ($\omega$).
This phonon density of state contains the vibrational information for a specific volume ($V$) and can be used to
calculate $F\lt{vib}$ at any temperature ($T$) with the Planck ($h$) and
Boltzmann constant (k\lt{b}) such that
\begin{eqnarray}
    G& = &E\lt{tot}(V)+F\lt{vib}(\omega(V),T)+pV \label{eq:G}\\
    F\lt{vib}& = &k\lt BT\int_\omega \mathrm{d}\omega D(\omega)\ln\left(2\sinh\left[\frac{h\omega}{4\pi k\lt BT}\right]\right)
\end{eqnarray}

To account for volume expansion during temperature increase the quasi harmonic approximation is used. 
For the diamond-like phases isotropic expansion is assumed.
The Gibbs energy is calculated for at least five different volumes.
The universal equation of state (EOS) \cite{VinetJoPCM1989} has been used to find the minimum
of $G$ with respect to the volume for a given temperature and pressure.
For all graphite-like phases and transition states anisotropic expansion along the c-axis has been included.
One parameter changes the unit cell isotropically and a second one changes only the c-axis.
A fourth order polynomial fit is used
to interpolate the EOS between the sampling points.
This increases the number of data points and also the accuracy.
Using the procedure described above we have calculated the Gibbs energies employing the
LDA for a wide range of pressures and temperatures.
However, due to the computational cost involved it is currently not possible to perform the same calculations
on the level of CCSD(T).
The CCSD(T) Gibbs energies have therefore been approximated using the following expression
\begin{equation}
    G^{\rm CCSD(T)}(V,T) \approx G^{\rm LDA}(V,T) - E^{\rm LDA}\lt{tot}(V_0)+ E^{\rm CCSD(T)}\lt{tot}(V_0) \label{eq:GCC},
\end{equation}
where $V_0$ corresponds to the LDA (equilibrium) volume of the (meta-)stable allotropes and the transition state
geometries. As such the volume  and temperature dependence of the CCSD(T) Gibbs energy is approximated
using LDA, which achieves sufficiently accurate descriptions of the phonons and bulk moduli for the purpose of
the present study.

\subsection{Phase transition probability}

\begin{figure*}[th]
  \centering
  \subfloat[$G\lt{\GABC}-G\lt{\pcTS}$]{\includegraphics[width=0.32\textwidth]{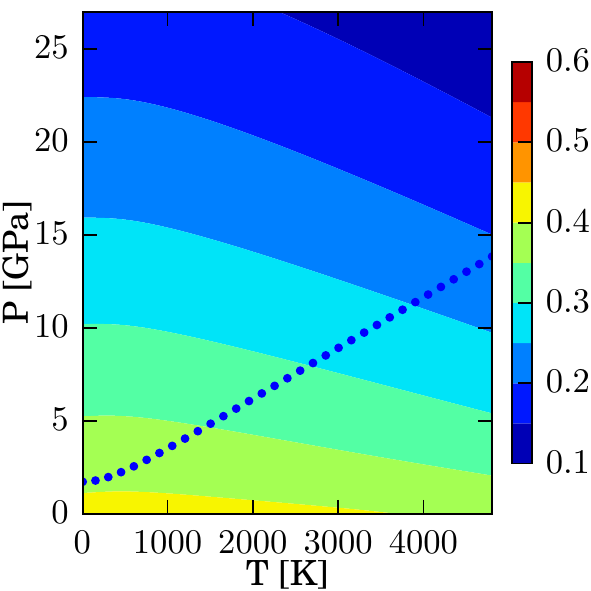}\label{fig:EAGpc}}~
  \subfloat[$G\lt{\GAB}-G\lt{\bwTS}$]{\includegraphics[width=0.32\textwidth]{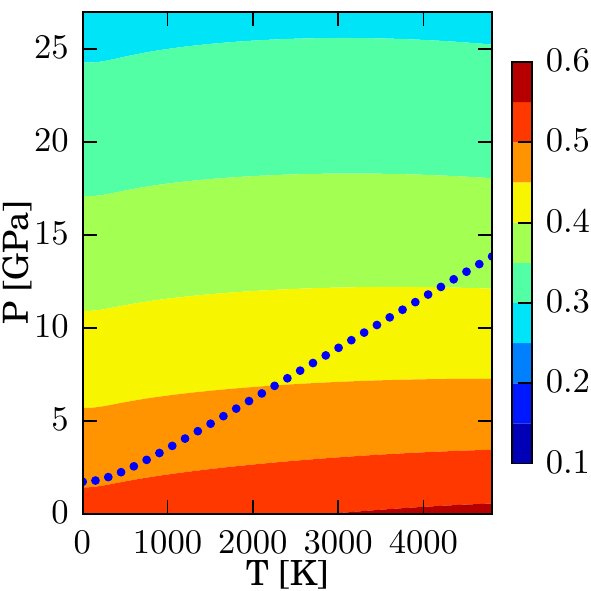}\label{fig:EAGbw}}~
  \subfloat[$G\lt{\GAB}-G\lt{\lpcTS}$]{\includegraphics[width=0.32\textwidth]{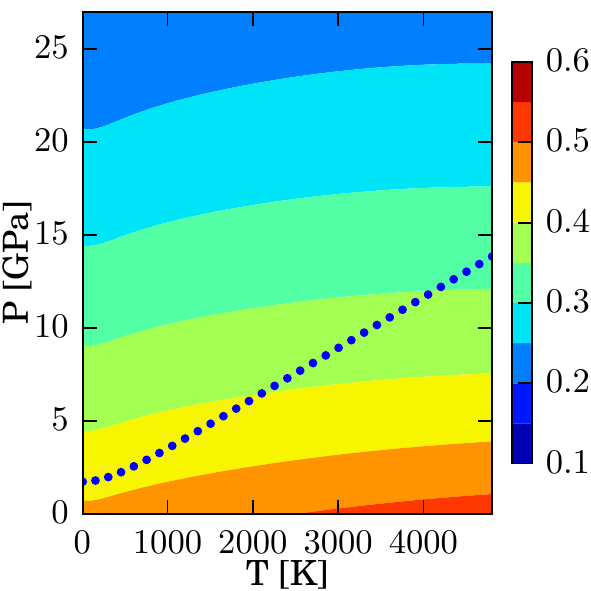}\label{fig:EAGlpc}}\\
  \subfloat[$f$ (\GABC~\ce{->[\pcTS]}~\cD)]{\includegraphics[width=0.32\textwidth]{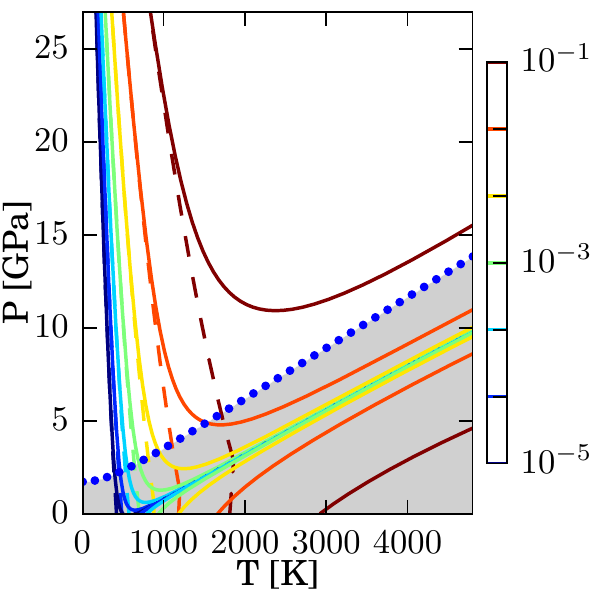}\label{fig:rateC}}~
  \subfloat[$G\lt{\hD}-G\lt{\lpcTS}$]{\includegraphics[width=0.32\textwidth]{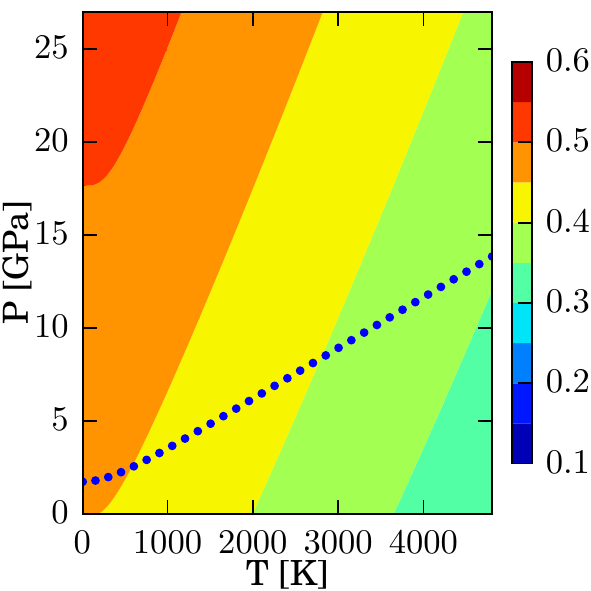}\label{fig:EADlpc}}~
  \subfloat[$f$ (\hD~\ce{->[\lpcTS]}~\cD)]{\includegraphics[width=0.32\textwidth]{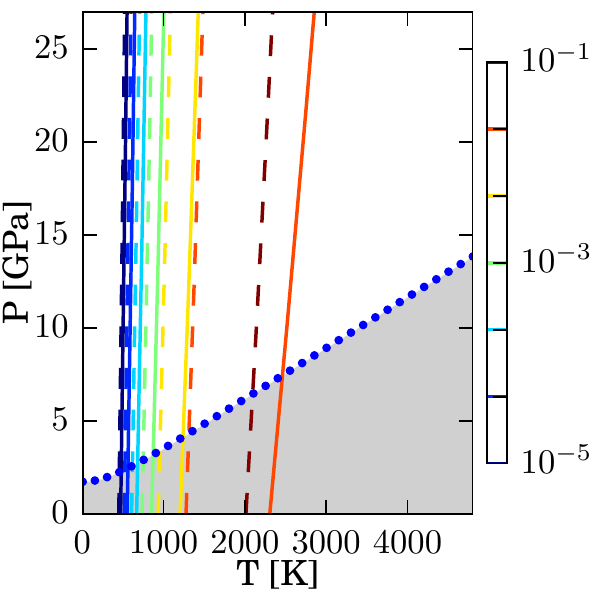}\label{fig:rateWCC}}
  \caption{Gibbs energy differences (activation energies) and probability ($f$) of phase transition for carbon.
  Starting with graphite \protect\subref{fig:EAGpc} produces \cD\ with the rate shown in \protect\subref{fig:rateC}.
  \protect\subref{fig:EAGbw} and \protect\subref{fig:EAGlpc} depict the activation energies to produce
  \hD\ and \cD,  respectively. \protect\subref{fig:EADlpc} refers
  to the transition from \hD\ to \cD\ with the rate shown in \protect\subref{fig:rateWCC}.
  The solid lines in \protect\subref{fig:rateC} and \protect\subref{fig:rateWCC} refer to \eref{eq:rate}, whereas the dashed lines excludes the backward reaction (second term in \eref{eq:rate}).}
\end{figure*}

In this work we will approximate the
probability that a phase transition occurs
using the activation energy only and disregard kinetic effects.
The activation energy ($\Delta G$) is the difference between the Gibbs energies of the transition
and initial states.
The Gibbs energy can be calculated as described in the previous subsection.
Close to the equilibrium phase boundary the back and forward reaction has to be taken
into account and the probability $f$
depends exponentially on $\Delta G$ 
\begin{equation}
 f = \exp(\frac{-\Delta G\lt{forward}}{RT}) - \exp(\frac{-\Delta G\lt{back}}{RT}) \label{eq:rate}
\end{equation}
with $R$ as the gas constant.
A comparable ansatz was published in Refs.~\cite{YafeiC1994,WangJoPCM1999}.

\fref{fig:EAGpc} to \ref{fig:rateWCC} shows the calculated CCSD(T) phase transition probabilities
and activation energies for carbon and a selection of transition states.
We note that the behavior of the transition probabilities
at low temperatures mostly arises from the explicit dependence of $f$ on the temperature in
the exponent rather than the temperature dependence of $\Delta G\lt{forward}$ or $\Delta G\lt{back}$.
Sec.~\ref{sec:phasediagramm} provides a more detailed discussion of the obtained results.

\section{Results and discussion}\label{sec:results}

We now turn to the discussion of the obtained DFT, HF and post-HF results.
The following section is organized as follows.
We first summarize the energy differences obtained using different methods for
carbon (sec.~\ref{sec:carbon}) and boron nitride (sec.~\ref{sec:bn})
allotropes, respectively.
Subsequently a comparison between results obtained for carbon and boron nitride will be drawn in sec.~\ref{sec:compCBN}. Sec.~\ref{sec:phasediagramm} employs the calculated ground state energies on the level of CCSD(T) theory
and the DFT results to predict the pressure-temperature phase diagrams of carbon and boron nitride.
In sec.~\ref{sec:revision} we review experimentally observed phase transitions and compare to the produced theoretical results.
Section \ref{sec:lonsdaleite} focuses on the hexagonal form of diamond.

\subsection{Carbon allotropes}\label{sec:carbon}

\begin{figure}[h]
 \includegraphics[]{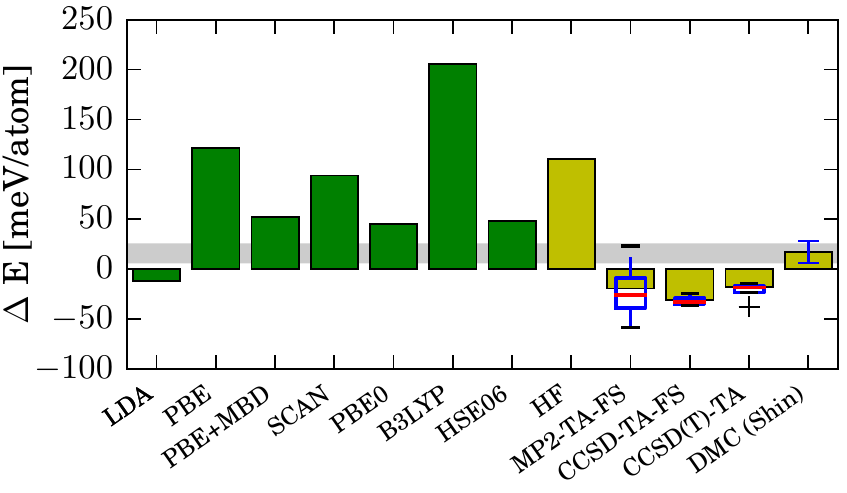}
 \caption{\label{fig:DGmethod}Energy difference between \cD\ and \GABC\ ($\Delta E=E_{\cD}-E_{\GABC}$) compared with DMC \cite{ShinTJoCP2014} and experimental values with a freely chosen error of \SI{+-10}{\meV\per\atom} (gray bar) \cite{WagmanJoRotNBoS1945}
excluding \SI{9}{\meV\per\atom} ZPVE.
 See text for further details.}
\end{figure}

\fref{fig:DGmethod} depicts the electronic ground state energy differences ($\Delta E=E_{\cD}-E_{\GABC}$)
between  carbon diamond (\cD) and graphite (\GABC) obtained using a range of DFT and quantum
chemical wavefunction based theories. Experimentally \GAB\ is the most
stable form of graphite. However, \GABC\ and \GAB\ are degenerate to within a few
meV per atom.
Positive and negative energy differences in \fref{fig:DGmethod} indicate the thermodynamic stability of graphite and diamond,
respectively. We stress that these calculations employ the DFT-LDA relaxed structures and
that further relaxation effects of the respective functionals are not taken into account.
Further relaxation effects can be significant for functionals that fail to describe the
interlayer binding in graphite. However, this section will focus on benchmarking the accuracy of the employed
functionals for a fixed geometry only.
In passing we note, however, that for comparison we have repeated the DFT calculations summarized
in \fref{fig:DGmethod} using geometries relaxed on the level of the PBE+MBD functional.
The corresponding energy differences did not change by more than five percent as a result of the small changes
in the employed geometries.

The grey bar in \fref{fig:DGmethod} corresponds to the experimental estimate of the ground state energy difference corrected for
zero-point vibrational energies (ZPVEs).
The experimental value of the difference in the Gibbs energy between graphite and diamond
is \SI{25}{\meV\per\atom} and has been obtained from the heat of combustion
and extrapolation to \SI{0}{\K} using the heat capacity \cite{WagmanJoRotNBoS1945}.
Therefore the latter value includes ZPVEs that
stabilize graphite compared to diamond.
To allow for a direct comparison between experiment and theory we have removed \SI{9}{\meV\per\atom}
ZPVE contributions (estimated using DFT-LDA) from the experimental energy difference.

We now turn to the discussion of the energy differences in \fref{fig:DGmethod} obtained using XC functionals
in the framework of DFT.
LDA underestimates the energy difference, predicting diamond to be more stable than
graphite by \SI{12}{\meV\per\atom}.
On the level of the GGA using the PBE functional we find that the stability of graphite is significantly
overestimated by almost \SI{100}{\meV\per\atom}  compared to experiment. This overestimation is partly
reduced by including dispersion effects on the level of MBD or by switching to the SCAN functional,
yielding energy differences of \SI{52}{\meV\per\atom} (PBE+MBD) and \SI{94}{\meV\per\atom} (SCAN), respectively.
Furthermore hybrid functionals such as PBE0 or HSE06 constitute a further
improvement compared to SCAN, overestimating the stability of graphite compared to experiment
by a few \SI{10}{\meV\per\atom} only. However, we note that the B3LYP hybrid functional does not follow
this trend and gives the worst agreement with experiment out of all theories considered in the present study.
Therefore a systematic improvability of the employed XC functionals with respect to their rung and computational cost
can not be achieved in the present case.
Furthermore one conclusion of the above findings is that non-van der Waals corrected
higher-level functionals
(PBE, SCAN, PBE0 and HSE06) predict the graphitic phase to be more stable than diamond,
whereas the inclusion of van der Waals corrections (MBD) can reverse their ordering.
Indeed we note in passing that in contrast to HSE06, HSE06+MBD predicts \cD~being more stable
than \GABC~by \SI{30}{\meV\per\atom}.

We now turn to the discussion of the results obtained using wavefunction based theories as depicted in
\fref{fig:DGmethod}. 
HF predicts graphite to be more stable than diamond by approximately \SI{100}{\meV\per\atom}, albeit
neglecting dispersion effects that play an important role in the interlayer binding of graphite.
We note that due to the neglect of these contributions, HF would predict the isolated graphene sheets to be more
stable than graphite. 
Second-order M\o{}ller-Plesset (MP2) perturbation theory corresponds to the next level of
wavefunction based method and predicts diamond to be slightly more stable than graphite by
approximately \SI{19}{\meV\per\atom}.
However, the finite size errors of the obtained MP2 results are significant as indicated
by the box plot, which is described in Sec.~\ref{sec:fseffects}.
The $k$-point mesh convergence using CCSD-TA-FS theory is much faster compared to MP2-TA-FS theory as indicated
by the smaller error bar.
We find that CCSD-TA-FS theory predicts diamond to be more stable than
graphite by \SI{31}{\meV\per\atom}.
Including the perturbative triples contribution to CCSD-TA-FS theory yields an even better agreement with
experiment albeit predicting diamond to be slightly more stable than graphite by \SI{14}{\meV\per\atom}.
In passing we note that DMC has been used in
Ref.~\cite{ShinTJoCP2014} to predict an energy difference in almost perfect agreement with experiment,
whereas the random-phase approximation (RPA) predicts both allotropes to be exactly degenerate~\cite{Lebegue2010}.
The good agreement between DMC and experiment is partly fortuitous due to remaining errors from the
stochastic sampling, the fixed-node approximation
and the employed pseudo-potentials.
However, the agreement between the high-level methods such as wavefunction based theories DMC and CCSD(T),
the RPA results and experiment to within a few ten \si{\meV\per\atom} is encouraging.
The remaining finite size and basis set errors in CCSD(T) theory calculations do not allow for
predicting which carbon allotrope is more stable, although we can conclude that they are expected to be degenerate
to about 10--20 \si{\meV\per\atom} including ZPVE.
Our findings indicate that quantum chemical wavefunction based theories allow for a systematic improvability of the
predicted energy differences as one increases the level of theory ranging from HF, MP2, CCSD to CCSD(T).

\begin{figure*}[t]
 \includegraphics[]{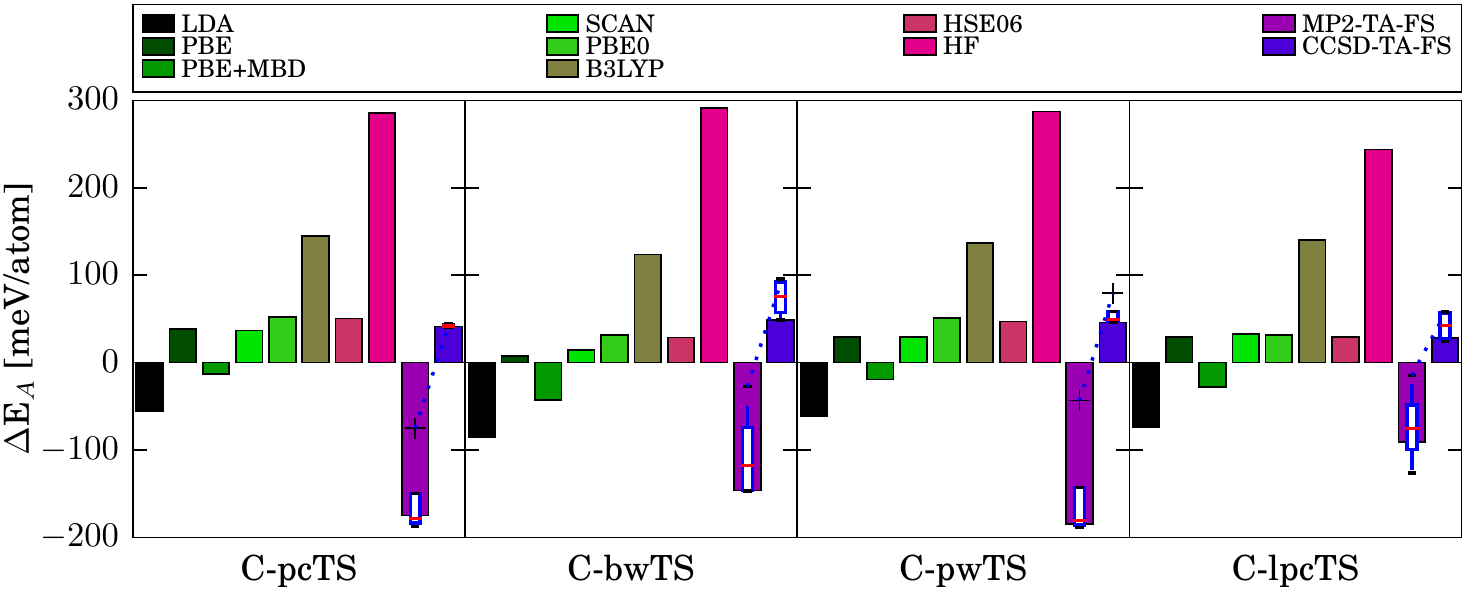}
 \caption{\label{fig:CTSGmethod}Energy difference between \GABC\ and transition state (TS) compared to CCSD(T)-TA referred as
 \SI{0}{\meV\per\atom}. Outlier marked with a ''+`` and dotted line connects energies from $k$-meshes with only one $k$-point along
 one direction. See text for further details.}
\end{figure*}
Having demonstrated that CCSD(T) theory is expected to yield accurate energy differences for
the thermodynamically most stable carbon allotropes we now seek to investigate the pressure-driven transition
pathways introduced in Sec.~\ref{sec:structurespt}.
To this end we focus on the activation barrier height that is defined as the difference in the electronic
ground state energy between graphite and the corresponding transition state $E_A=E_{\rm TS}-E_{\rm \GABC}$.
The considered transition states are referred
to as \pcTS, \pwTS, \bwTS\ and \lpcTS.
The activation barriers can not be compared to experimental observations directly but serve as theoretical
benchmark systems and qualitative models for realistic phase transitions.
\fref{fig:CTSGmethod} depicts the difference of calculated activation barrier heights between various methods
and CCSD(T) (including finite size corrections); for example, $\Delta E_A^{\rm LDA}= E_A^{\rm LDA} - E_A^{\rm CCSD(T)-TA}$. 
The depicted results confirm well-known trends for the accuracy of DFT methods,
assuming that CCSD(T) theory can be considered an accurate benchmarking reference for the activation barrier height.
LDA underestimates the activation barrier heights for all investigated transition states by  \SIrange{50}{100}{\meV\per\atom},
showing that this level of theory suffers from larger errors in the description of XC energies for transition states compared to
initial and final states (\GABC\ and \cD).
Including gradient corrections on the level of the PBE functional improves the agreement
with CCSD(T) theory noticeably, yielding overestimated activation energies with errors smaller
than \SI{50}{\meV\per\atom} for all four transition states. This is in contrast to PBE results for
molecular activation barrier heights in the gas phase that are in general underestimated~\cite{doi:10.1021/cr200107z}.
However, we believe that the overestimation of the PBE barriers for the studied solids is caused
by the neglect of interatomic van der Waals forces, which play an important role for the present systems.
We stress that the inclusion of dispersion effects to PBE on the level of PBE+MBD theory
yields again underestimated activation energies that agree with CCSD(T) to within \SI{50}{\meV\per\atom}.
The SCAN functional yields barrier heights that are almost identical to
our PBE findings. Furthermore the inclusion of non-local exchange in the PBE0 and HSE06
hybrid functionals yields on average slightly larger barrier heights. We note that adding the MBD effect
(from the difference between PBE and PBE+MBD calculations) to these hybrid functionals would yield barrier
heights in almost perfect agreement with CCSD(T) theory.
On the other hand, we find that the B3LYP hybrid functional yields overestimated barrier heights, exhibiting
errors on a scale of more than \SI{100}{\meV\per\atom}.

We now turn to the discussion of activation barrier heights calculated using wavefunction based theories
starting with the HF method. Our findings are depicted in \fref{fig:CTSGmethod} and show that HF yields strongly
overestimated barrier heights with errors on the scale of almost \SI{300}{\meV\per\atom} compared to CCSD(T).
This trend is known from molecular quantum chemistry and can be explained
by the fact that HF neglects electronic correlation effects, which are in general larger in the transition
state compared to the initial and final state of most chemical reactions.
Accounting for electronic correlation effects on the level of MP2 theory corrects for this tendency although
it yields underestimated barriers by about \SIrange{100}{200}{\meV\per\atom} for all transition states.
We note that the box plots of the MP2 results in \fref{fig:CTSGmethod} also indicate that the remaining
finite size errors for these estimates are significant. We attribute this to the observation that some
transition states exhibit a metallic character in DFT calculations (\lpcTS\ and \bwTS)
and that MP2 theory suffers from severe shortcomings in metals such as $k$-point mesh divergence~\cite{PhysRevLett.110.226401}.
CCSD results for the barrier heights constitute a substantial improvement over MP2 findings,
exhibiting errors compared to CCSD(T) that are smaller than \SI{50}{\meV\per\atom}.
Furthermore we note that the box plot for CCSD-TA-FS results is significantly smaller, indicating
that the remaining finite size errors are below a few \SI{10}{\meV\per\atom}.
From these findings we conclude that quantum chemical wavefunction based theories including MP2 and CC
theories have the potential of achieving results for activation barrier heights in solid-solid phase
transitions with systematically improvable accuracy.

\begin{figure}[t]
 \includegraphics[]{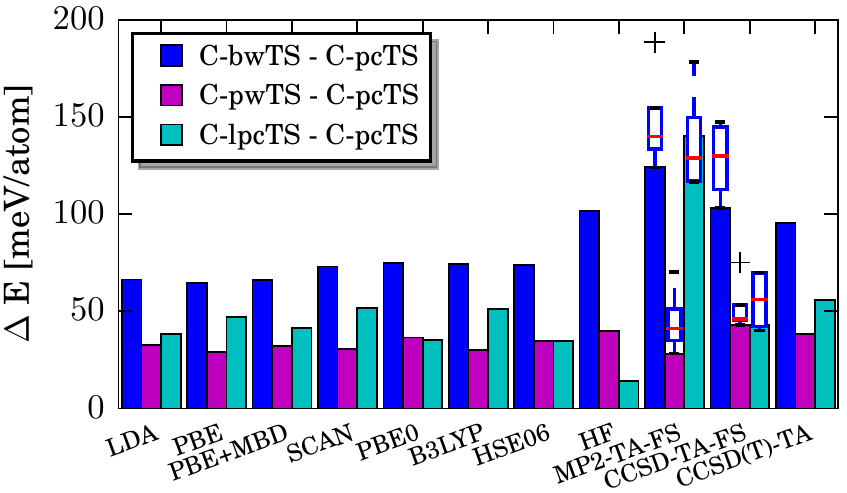}
 \caption{\label{fig:CTSpcmethod}Energy difference between \pcTS\ and other transition states.
 See text for further details.}
\end{figure}

Having discussed the accuracy of DFT and wavefunction based methods for predicting the activation barrier
heights, we now seek to address the question: which transition states are energetically the most favorable?
This is an important question because it affects through which transition state a pressure-driven phase transition
proceeds and which (meta-)stable carbon allotrope will be the outcome.
\fref{fig:CTSpcmethod} depicts the energy differences of the activation barrier heights with respect to the \pcTS\ for the respective electronic
structure theories.
Unequivocally all theories predict the \pcTS\ to be the energetically most favorable transition state,
implying that the puckering mechanism is expected to play the most important role in pressure-driven graphite to diamond transitions.
In passing we note that our LDA results are comparable to previous work~\cite{FahyPRB1987,TateyamaPRB1996}
and that the energy difference between the boat and chair conformation of graphane is \SI{55}{\meV\per\atom},
favoring the chair conformation \cite{SofoPRB2007}.
As regards the ordering of the remaining transition states (\bwTS, \pwTS, and \lpcTS), we find that the DFT methods
shown in \fref{fig:CTSpcmethod} predict all very similar orderings. \bwTS\ is energetically the least favorable transition state,
whereas \pwTS\ and \lpcTS\ agree to within a few meV per atom, except for the PBE, SCAN and
B3LYP functionals that predict the \pwTS\ to be slightly more favorable in energy than the \lpcTS.
In the case of results obtained using wavefunction based methods depicted in \fref{fig:CTSpcmethod} we find that
the \bwTS\ corresponds to the largest activation barrier height and that \pwTS\ and \lpcTS\ agree
to within the remaining finite size errors. However, MP2 theory deviates from this trend by predicting
equally large activation barrier heights for the \bwTS\ and \lpcTS, making the \pwTS\ the second most favorable transition state.
However, we stress that MP2 results are perhaps not meaningful due to
the metallic character of some transition states.
From the above results we conclude that interatomic van der Waals forces play a minor role in the
ordering of the respective transition states.
Furthermore the ordering is already correctly described on the level of the LDA to the XC functional.

\tref{tab:Cenergy} summarizes all energy differences discussed above for the (meta-)stable carbon allotropes and the transition states.
Furthermore the table also lists the energy difference between \cD\ and \hD\ that is predicted by all methods to be about \SI{30}{\meV\per\atom}.

\begin{table*}[ht]
  \begin{center}
    \footnotesize
    \caption{\label{tab:Cenergy} Energy differences in meV per atom for carbon structures as obtained by various levels of theory. }
    \begin{tabular}{ld{3.0}d{3.0}d{3.0}d{3.0}d{3.0}d{3.0}d{3.0}d{3.0}d{3.5}d{3.5}d{3.0}}
      \hline
      \hline
System	 & \multicolumn{1}{c}{LDA}	 & \multicolumn{1}{c}{PBE}	 & \multicolumn{1}{c}{PBE+MBD}	 & \multicolumn{1}{c}{SCAN}	 & \multicolumn{1}{c}{PBE0}	 & \multicolumn{1}{c}{B3LYP}	 & \multicolumn{1}{c}{HSE06}	 & \multicolumn{1}{c}{HF}	 & \multicolumn{1}{c}{MP2-TA-FS}	 & \multicolumn{1}{c}{CCSD-TA-FS}	 & \multicolumn{1}{c}{CCSD(T)-TA}\\
\hline
\cD\ – \GABC	 & -12	 & 121	 & 52	 & 94	 & 45	 & 206	 & 49	 & 110	 & -19 .\pm 29	 & -31 .\pm 5	 & -14\\
\hD\ -- \cD	 & 25	 & 24	 & 25	 & 29	 & 27	 & 30	 & 27	 & 37	 & 30 .\pm 14	 & 35 .\pm 10	 & 33\\
\pcTS\ -- \GABC 	 & 329	 & 423	 & 372	 & 422	 & 437	 & 530	 & 435	 & 671	 & 210 .\pm 46	 & 426 .\pm 1	 & 385\\
\bwTS\ -- \GABC 	 & 395	 & 488	 & 438	 & 495	 & 512	 & 604	 & 509	 & 772	 & 334 .\pm 49	 & 529 .\pm 20	 & 481\\
\pwTS\ -- \GABC 	 & 362	 & 452	 & 404	 & 452	 & 474	 & 560	 & 470	 & 710	 & 238 .\pm 61	 & 469 .\pm 14	 & 423\\
\lpcTS\ -- \GABC	 & 367	 & 470	 & 413	 & 473	 & 473	 & 581	 & 470	 & 685	 & 350 .\pm 41	 & 469 .\pm 16	 & 441\\
      \hline
      \hline
    \end{tabular}
  \end{center}
\end{table*}

\subsection{Boron nitride allotropes}\label{sec:bn}

\begin{figure}[t]
 \includegraphics[]{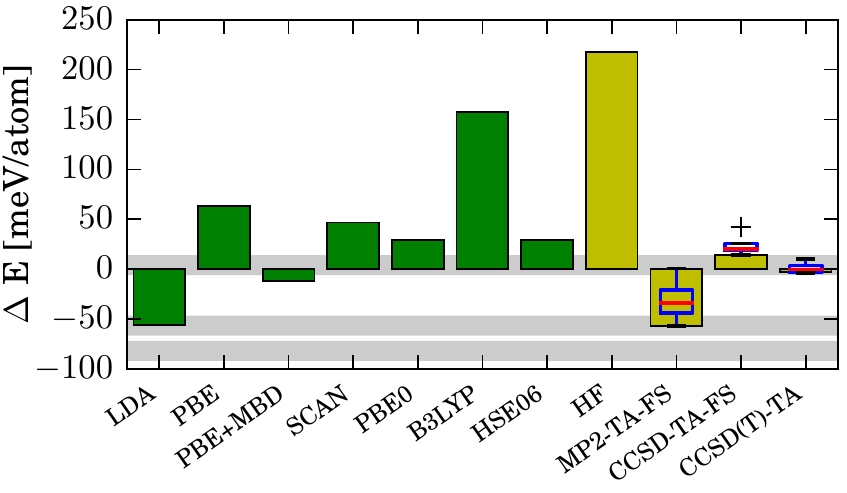}
 \caption{\label{fig:CRmethod}Energy difference ($\Delta E=E_{\cBN}-E_{\rBN}$) between \cBN\ and \rBN\ with
different methods compared to
 experimental values with a freely chosen error of \SI{+-10}{\meV\per\atom} (grey bar). 
 Experimental values refer to \hBN\ instead of \rBN\ extrapolated to \SI{0}{\K}: 
 \SI{-82}{\meV\per\atom} (Ref.\cite{SolozhenkoHPR1995}), \SI{-57}{\meV\per\atom} (Ref.~\cite{JeongJoNaN2013}), \SI{4}{\meV\per\atom} (Ref.~\cite{DayAM2012}) excluding \SI{4}{\meV\per\atom} ZPVE.}
\end{figure}

This section is organized similarly to Sec.~\ref{sec:carbon} and summarizes the boron nitride results.
\fref{fig:CRmethod} depicts the difference in the electronic ground state energies between \rBN\ and \cBN\ as obtained
using a range of DFT and wavefunction based methods, whereas experimental estimates refer to the difference between
\hBN\ and \cBN.
We stress that experimentally \hBN\ is always found to be more stable than \rBN.
However, calculations at zero pressure and temperature have shown that these two structures differ in
energy by less than \SI{4}{meV\per\atom} \cite{AlbePRB1997,ConstantinescuPRL2013}.
\rBN\ has a two atomic unit cell only, whereas \hBN\ contains at least four atoms in the unit cell.
Therefore calculations of \rBN\ are computationally less demanding.
We stress that all employed structures have been optimized using DFT-LDA.
In this section we focus on benchmarking the accuracy of the predicted energies for a fixed geometry.
In passing we note, however, that for comparison we have repeated the DFT calculations summarized
in \fref{fig:CRmethod} using geometries relaxed on the level of the PBE+MBD functional.
The corresponding energy differences did not change by more than five percent as a result of the small changes
in the employed geometries.

Positive and negative energy differences in \fref{fig:CRmethod} indicate that the low- (\hBN / \rBN) and high-density (\cBN) phase
is predicted to be more stable, respectively.
Grey bars show experimental findings.
Solozhenko et al. predict \cBN\ to be more stable than \hBN\ by \SI{78}{\meV\per\atom} at \SI{0}{\K} (including ZPVE) \cite{SolozhenkoHPR1995}.
This result was obtained from fluorine combustion and extrapolation to \SI{0}{\K} using the heat capacity.
Other recent experiments obtain the equilibrium phase boundary directly from catalytic transitions
with X-ray diffraction analysis \cite{FukunagaDaRM2000}, finding that \hBN\ is more stable even at \SI{0}{\K}.
To allow for a comparison between experiment and theory we have removed \SI{4}{\meV\per\atom} ZPVE from the 
experimental estimates in \fref{fig:CRmethod} estimated using LDA.

We find that DFT-LDA calculations predict \cBN\ to be more stable than \rBN\ with an energy difference of \SI{-56}{\meV\per\atom}
in good agreement with results from Ref.~\cite{KernPRB1999}.
Including gradient corrections on the level of the PBE functional reverses their order and yields an energy
difference of \SI{64}{\meV\per\atom}. However, it is known that van der Waals interactions have to be taken into account for an accurate
description of electronic correlation effects especially in layered compounds.
Inclusion of MBD on top of the PBE functional allows for capturing such correlation effects. Compared to PBE, PBE+MBD reverses the order between
both allotropes again, yielding an energy difference of \SI{-13}{\meV\per\atom}.
The SCAN functional goes beyond the GGA and is expected to perform better than PBE~\cite{PhysRevLett.115.036402}.
Our findings shown in \fref{fig:CRmethod} reveal that SCAN reduces the energy difference between \cBN\ and \rBN\ compared
to PBE from \SI{64}{\meV\per\atom} to \SI{47}{\meV\per\atom}.
The hybrid functionals PBE0 and HSE06 continue this trend and predict \rBN\ to be more stable than \cBN\ 
with an even smaller energy difference of \SI{30}{\meV\per\atom} and \SI{31}{\meV\per\atom}, respectively. However, the B3LYP functional
significantly overestimates the stability of \rBN, predicting a difference of \SI{160}{\meV\per\atom}.
The different results from the various DFT methods make it difficult to provide a firm conclusion on the true energy difference.
However, we believe that the results allow for a similar conclusion as for the case of carbon
allotropes: higher-level functionals (PBE, SCAN, PBE0 and HSE06) predict the graphitic phase to be more
stable than the diamond-like phase but the inclusion of van der Waals corrections can reverse their ordering.

We now turn to the discussion of results for the energy difference between \cBN\ and \rBN\ 
obtained using wavefunction based methods as depicted in \fref{fig:CRmethod}. 
HF theory, disregarding electronic correlation effects, substantially overestimates the stability of \rBN\ compared to \cBN.
The HF energy difference is the largest of all considered methods (\SI{222}{\meV\per\atom}). The simplest treatment of electronic correlation effects
on the level of wavefunction based methods is achieved using MP2 theory, predicting \cBN\ to be more stable than \rBN\ by
\SI{58}{\meV\per\atom}. However, as indicated by the large error bars we find that the remaining finite size errors on the level
of MP2 theory are on the scale of several \SI{10}{\meV\per\atom}.
CCSD-TA-FS predicts \rBN\ to be more stable than \cBN\ by \SI{14}{\meV\per\atom} and is well converged with respect to the
employed $k$-mesh.
The inclusion of perturbative triples yields an energy difference of \SI{2}{\meV\per\atom} only.
From these findings we conclude that the series of wavefunction based theories ranging from HF, MP2, CCSD to CCSD(T) predicts
an oscillating but convergent energy difference between  \rBN\ and \cBN\ that is close to zero on
the level of CCSD(T) theory.
Due to the remaining finite size errors that are estimated to be on the scale of approximately \SI{10}{\meV\per\atom}, we conclude
that \rBN\ and \cBN\ are degenerate to within \SI{10}{\meV\per\atom} on the level of CCSD(T).

\begin{figure*}[ht]
 \includegraphics[]{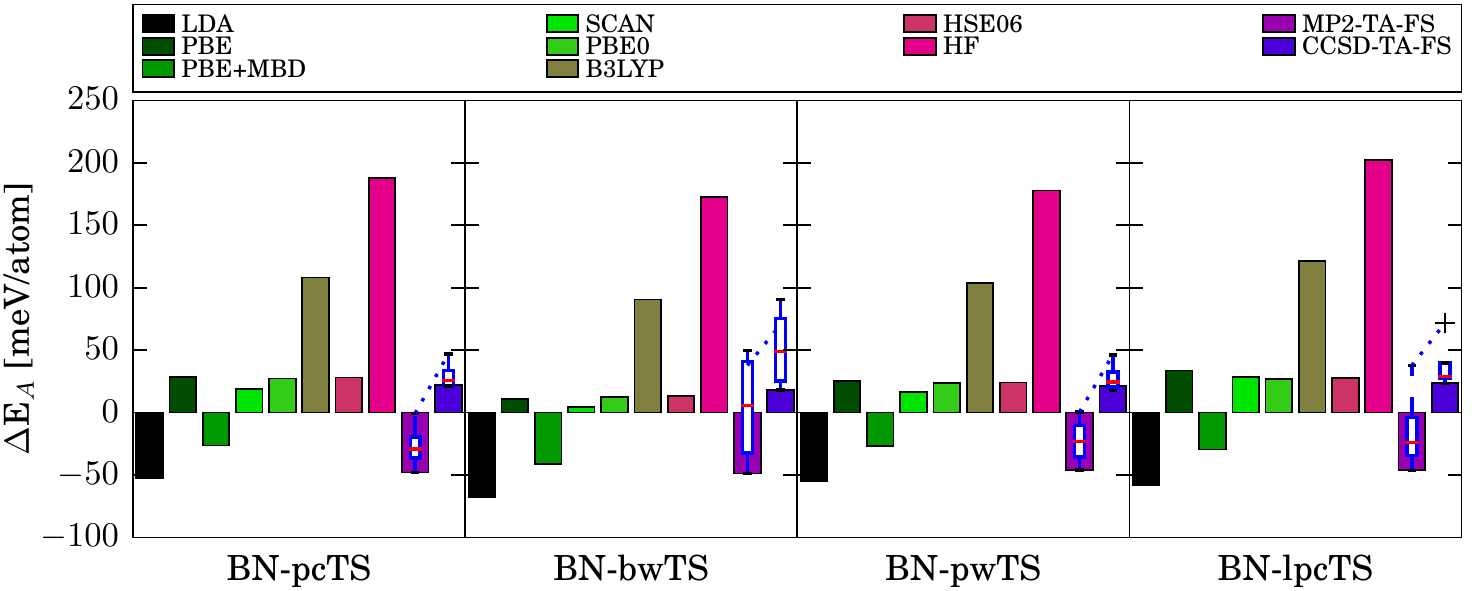}
 \caption{\label{fig:BNTSGmethod}Energy difference between \hBN\ and transition state (TS) compared to CCSD(T)-TA referred as \SI{0}{\meV\per\atom}.
 Outlier marked with a ''+`` and dotted line connects energies with the smallest $k$-meshes.
 See text for further details.}
\end{figure*}

We now seek to investigate the pressure-driven transition pathways introduced in Sec.~\ref{sec:structurespt} for boron nitride.
The discussion and most results are analogue to our findings for carbon.
We investigate again the activation barrier height that is defined as the difference in the
electronic ground state energy between \hBN\ and the corresponding transition state $E_A=E_{\rm TS}-E_{\hBN}$.
\fref{fig:BNTSGmethod} depicts the difference in activation barrier heights
between various methods and CCSD(T) (including finite 
size corrections); for example, $\Delta E_A^{\rm LDA}= E_A^{\rm LDA} - E_A^{\rm CCSD(T)-TA}$. 
LDA underestimates all activation barrier heights by approximately \SI{50}{\meV\per\atom}.
Including the effect of gradient corrections on the level of the PBE functional improves the agreement
with CCSD(T) theory slightly, yielding overestimated activation energies with average errors of roughly \SI{25}{\meV\per\atom} for
all four transition states. This is in contrast to PBE results for molecular activation barrier heights in the gas phase that
are in general underestimated.
However, we believe that the overestimation of the PBE barriers for the studied solids is caused
by the neglect of interatomic van der Waals forces in the same manner as for carbon.
We stress that the inclusion of dispersion effects to PBE on the level of PBE+MBD theory
yields again underestimated activation energies that agree with CCSD(T) to within approximately \SI{25}{\meV\per\atom}.
Moving to the next level of theory, we find that the SCAN functional yields barrier height results that are slightly better than
our PBE findings.
The PBE0 and HSE06 hybrid functionals yield barrier heights similar to SCAN. 
However, the B3LYP hybrid functional overestimates barrier heights substantially, exhibiting errors on a scale
of more than \SI{100}{\meV\per\atom}.

We now turn to the discussion of activation barrier heights calculated on the level of wavefunction based theories,
starting with the HF method.
Our findings are depicted in \fref{fig:BNTSGmethod} and show that HF yields significantly overestimated barrier heights with errors
ranging from \SI{150}{\meV\per\atom} to \SI{200}{\meV\per\atom} compared to CCSD(T).
Accounting for electronic correlation effects on the level of MP2 theory corrects for this tendency partly
despite yielding significantly underestimated barriers on the scale of \SI{50}{\meV\per\atom} and suffering from non-negligible finite 
size errors as indicated by the error bars.
CCSD results for the barrier heights constitute a substantial improvement over MP2 findings, 
overestimating the barriers by about \SI{25}{\meV\per\atom}.
From these findings we conclude again that quantum chemical wavefunction based theories including MP2 and CC theories have the potential of achieving
results for activation barrier heights in solid-solid phase transitions with systematically improvable accuracy.
However, their finite size errors
are a dominant source of error in our present calculations.

\begin{figure}[h]
 \includegraphics[]{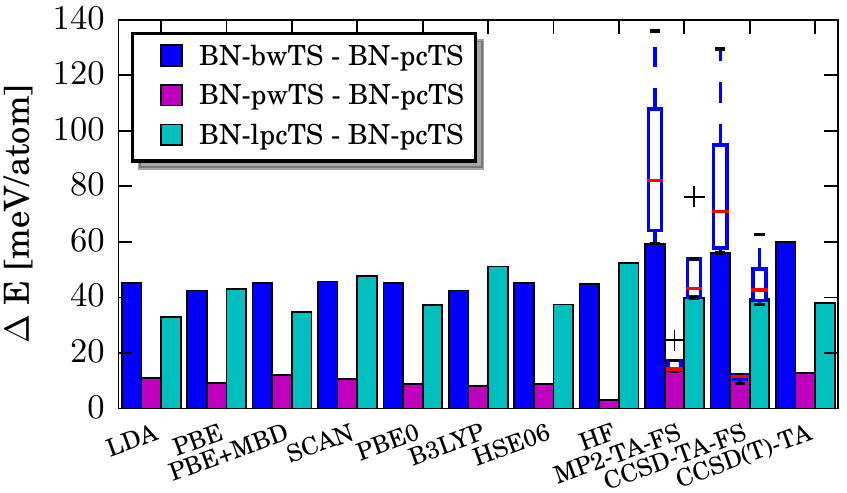}
 \caption{\label{fig:BNTSpcmethod}Energy difference between \pcTS\ and other transition states.
 See text for further details.}
\end{figure}
\begin{table*}[ht]
  \begin{center}
    \footnotesize
    \caption{\label{tab:BNenergy}  Energy differences in meV per atom for boron nitride structures as obtained by various levels of theory.}
    \begin{tabular}{ld{3.0}d{3.0}d{3.0}d{3.0}d{3.0}d{3.0}d{3.0}d{3.0}d{3.5}d{3.5}d{3.0}}   
\hline \hline
System	 & \multicolumn{1}{c}{LDA}	 & \multicolumn{1}{c}{PBE}	 & \multicolumn{1}{c}{PBE+MBD}	 & \multicolumn{1}{c}{SCAN}	 & \multicolumn{1}{c}{PBE0}	 & \multicolumn{1}{c}{B3LYP}	 & \multicolumn{1}{c}{HSE06}	 & \multicolumn{1}{c}{HF}	 & \multicolumn{1}{c}{MP2-TA-FS}	 & \multicolumn{1}{c}{CCSD-TA-FS}	 & \multicolumn{1}{c}{CCSD(T)-TA} \\
\hline 
\cBN\ -- \rBN	 & -56	 & 64	 & -13	 & 47	 & 30	 & 160	 & 31	 & 222	 & -58 .\pm 21	 & 14 .\pm 11	 & 2\\
\wBN\ -- \cBN	 & 18	 & 17	 & 18	 & 20	 & 19	 & 20	 & 19	 & 23	 & 23 .\pm 7	 & 24 .\pm 4	 & 22\\
\pcTS\ -- \hBN	 & 169	 & 250	 & 195	 & 240	 & 249	 & 329	 & 249	 & 409	 & 173 .\pm 17	 & 243 .\pm 10	 & 221\\
\bwTS\ -- \hBN	 & 214	 & 292	 & 240	 & 286	 & 294	 & 372	 & 294	 & 454	 & 233 .\pm 42	 & 299 .\pm 30	 & 281\\
\pwTS\ -- \hBN	 & 179	 & 259	 & 207	 & 251	 & 258	 & 338	 & 258	 & 412	 & 188 .\pm 18	 & 256 .\pm 11	 & 234\\
\lpcTS\ -- \hBN	 & 202	 & 293	 & 230	 & 288	 & 286	 & 381	 & 287	 & 461	 & 213 .\pm 32	 & 283 .\pm 19	 & 259\\
\hline \hline
    \end{tabular}
  \end{center}
\end{table*}

Having discussed the accuracy of DFT and wavefunction based methods for predicting the activation barrier heights, we now seek to
address the question which transition states are energetically the most favorable for boron nitride allotropes.
As for carbon this is an important question because it affects through which transition state a pressure-driven phase
transition proceeds and which (meta-)stable boron nitride allotrope will be the outcome.
\fref{fig:BNTSpcmethod} depicts the energy differences of the activation barrier heights with respect to the \pcTS\ for all employed electronic
structure theories.
Unequivocally all theories predict the \pcTS\ to be the energetically most favorable transition state in the same manner as for carbon.
As regards the ordering of the remaining transition states (\bwTS, \pwTS, and \lpcTS), we find that all methods
predict \pwTS\ to be the second most favorable transition state, whereas \bwTS\ and \lpcTS\ have the largest barrier heights.
An important conclusion from the above results is that all levels of theory agree qualitatively and predict similar energy differences
between the barrier heights. However, we note in passing that our MP2 and CCSD results exhibit very large finite size errors for the \bwTS.

All the energies discussed in this section are summarized in \tref{tab:BNenergy}.
Furthermore we note that all employed methods predict \wBN\ to be less stable than \cBN\ by about \SI{20}{\meV\per\atom}.

\subsection{Comparing the carbon and boron nitride systems}\label{sec:compCBN}

As discussed in the previous sections and as summarized in \tref{tab:Cenergy} and \ref{tab:BNenergy}, the produced
results for carbon and boron nitride systems are very similar.
Overall the employed electronic structure theories exhibit the same trends for the prediction of
energy differences between (meta-)stable allotropes and barrier heights.
However, one interesting exception is the energy difference between the high- (\cBN/\cD) and low-density (\rBN/\GABC) phases.
Comparing this difference for DFT based methods between boron nitride and carbon systems reveals that DFT tends to stabilize the high-density phases
of boron nitride more than in the case of carbon.
The opposite trend can be observed for wavefunction based methods with the exception of MP2 theory.
This trend is most evident when comparing LDA (BN: \cBN $-$ \rBN = \SI{-56}{\meV\per\atom}; C: \cD$-$\GABC =  \SI{-12}{\meV\per\atom})
to HF (BN: \cBN $-$ \rBN = \SI{222}{\meV\per\atom}; C: \cD$-$\GABC =  \SI{110}{\meV\per\atom}).
Improving upon the respective rung of theory makes this trend less pronounced as can be seen for
HSE06 (BN: \cBN $-$ \rBN = \SI{31}{\meV\per\atom}; C: \cD$-$\GABC =  \SI{49}{\meV\per\atom})
and CCSD(T) (BN: \cBN $-$ \rBN = \SI{2}{\meV\per\atom}; C: \cD$-$\GABC =  \SI{-14}{\meV\per\atom}).
This shows that the true energy differences between low- and high-density phases of carbon and boron nitride systems
are perhaps on the scale of ten meV per atom, indicating that their respective phase diagrams are very similar.

Moreover we note that the barrier heights in carbon are approximately
\SI{180}{\meV\per\atom} larger than for boron nitride which can be attributed to the stronger
covalent bond in carbon allotropes.

\subsection{Phase diagrams and transition probabilities}\label{sec:phasediagramm}

\begin{figure*}[ht]
  \centering
  \subfloat[]{\label{fig:phasediagramC}\includegraphics[]{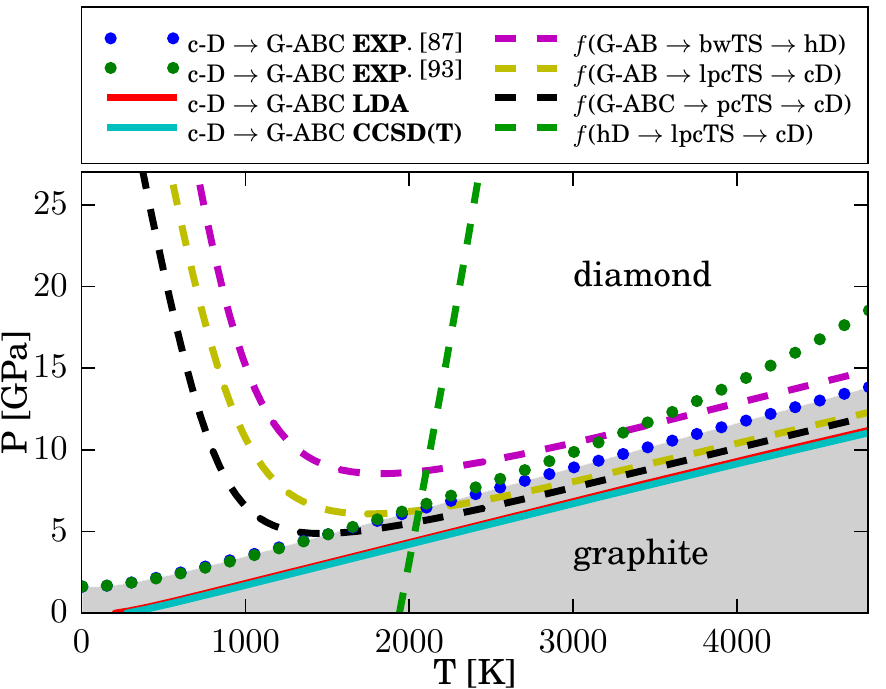}}
  \subfloat[]{\label{fig:phasediagramBN}\includegraphics[]{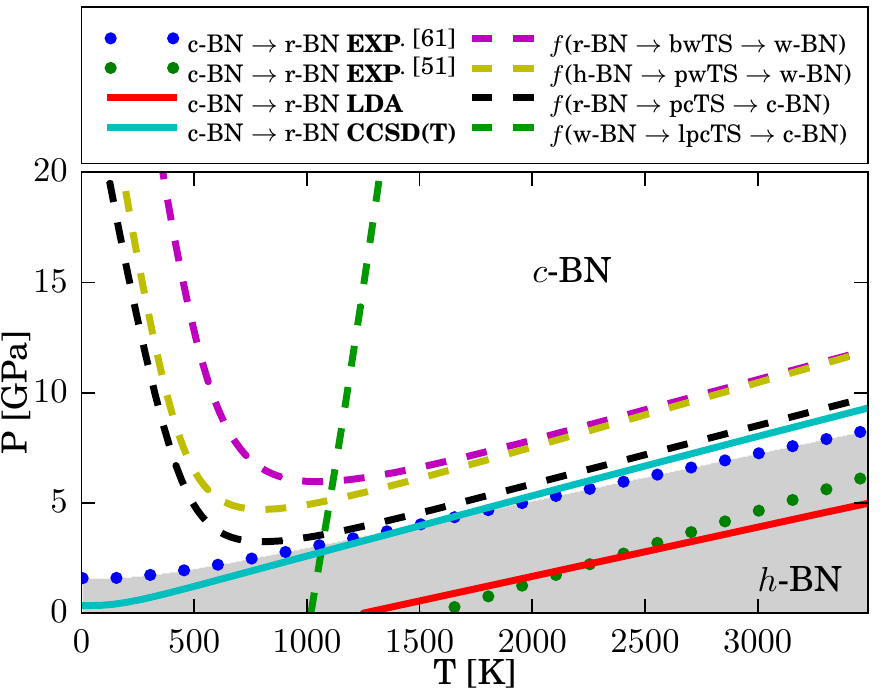}}
  \caption{\label{fig:phasediagram} Pressure-temperature phase diagrams for carbon (a) and boron nitride (b).
  The calculated  LDA and CCSD(T) equilibrium phase boundaries are depicted by solid lines.
  The dotted lines correspond to the experimental equlibrium phase boundaries as obtained from the Berman-Simon
  line \cite{YafeiC1994} and Day \etal\ \cite{DayAM2012} for carbon
  and from Fukunaga \etal\ \cite{FukunagaDaRM2000} and Solozhenko \etal\ \cite{SolozhenkoHPR1995} for boron nitride.
  The phase tranistion probability contour lines for a fixed probability of $f = 0.02$ are shown by dashed lines.
}
\end{figure*}

We now turn to the discussion of pressure-temperature phase diagrams for carbon and boron nitride.
The calculated phase diagrams are obtained from the 
Gibbs energies ($G$) as defined in \eref{eq:G}.
Furthermore the Gibbs energy of the transition states $G\lt{TS}(T,P)$ allows for determining
approximate phase transformation probabilities as defined by \eref{eq:rate}.

\fref{fig:phasediagram} depicts the pressure-temperature phase diagrams for carbon and boron nitride.
The experimental equilibrium phase boundaries are shown by dotted lines and separate the low-density
graphitic phases (\hBN\ and \GAB) at lower pressures and higher temperatures from the high-density
diamond-like phases (\cBN\ and \cD) at higher pressures and lower temperatures.
In the case of carbon, the two experimental phase boundaries deviate from each other only at
temperatures higher than \SI{2000}{\K}~\cite{YafeiC1994,DayAM2012}. Furthermore
both experimental results predict graphite being the stable carbon allotrope at ambient conditions.
This is in contrast to boron nitride where the experimental equilibrium phase boundaries disagree
by several GPa at lower temperatures~\cite{SolozhenkoHPR1995,FukunagaDaRM2000}.
The experimental findings by Solozhenko \etal\ \cite{SolozhenkoHPR1995} 
and Fukunaga \etal\ \cite{FukunagaDaRM2000} even differ in their prediction of
the thermodynamically most stable allotrope at ambient conditions.
Solozhenko \etal\ predict the zinc blende phase, whereas Fukunaga \etal\ find the \hBN\ phase being more stable.

\fref{fig:phasediagram} also shows the calculated equilibrium phase boundaries obtained using DFT
on the level of the LDA. In the case of carbon, the LDA results agree with
both experimental findings to within a few GPa at low temperatures. At temperatures higher than
\SI{2000}{\K} the LDA boundary is closer to the Berman-Simons line.
However, we stress that LDA results are less accurate for higher temperatures due
to the underestimation of the interlayer binding energy.
In the case of boron nitride, we find that
the LDA phase boundary agrees well with that of
Solozhenko \etal\ \cite{SolozhenkoHPR1995},
predicting that the high-density phase (\cBN) is more stable than the low-density (\hBN)
phase at ambient conditions.
However, this agreement is most likely fortuitous
due to the limited accuracy of the LDA as discussed in Secs.~\ref{sec:carbon},\ref{sec:bn} and \ref{sec:compCBN}.
\fref{fig:phasediagram} also depicts the equilibrium phase boundaries
obtained using CCSD(T) theory.
We stress that the employed CCSD(T) Gibbs energies are approximated using Eq.\eqref{eq:GCC},
corresponding to a rigid shift of the LDA Gibbs energies.
Compared to the LDA, the more accurate CCSD(T) theory shifts the equilibrium phase boundary
of carbon only slightly (\SI{0.2}{\GPa}), whereas it has a large effect on the phase boundary of boron nitride.
We find that CCSD(T) theory predicts
the \hBN\ phase being slightly more stable than \cBN\ at ambient conditions in good agreement with
the boundary of Fukunaga \etal\ \cite{FukunagaDaRM2000}.
We atribute the disagreement between the experimental and calculated phase boundaries at
higher temperatures to the underestimation of the interlayer binding in \hBN\ on the level of the LDA
and the neglect of anharmonic vibrational energy contributions~\cite{KernPRB1999}.
In passing we note that the equilibrium phase boundary of carbon allotropes has also been investigated
using HSE06, GGA and GGA+vdW functionals in Ref.\cite{Grochala2014-ij,YU2014185}.

Atomistic simulations of the pressure- and temperature-driven phase transitions in the considered carbon and boron
nitride allotropes are computationally demanding even on the level of DFT \cite{ScandoloPRL1995,KhaliullinNM2011}.
However, it is possible to gain insight into the required pressure and temperature conditions of
phase transitions from a minimal model using the considered (meta-)stable and
transition states, their Gibbs energies and an approximate expression for the transition probability
given by \eref{eq:rate}~\cite{YafeiC1994}. 
We have performed calculations of these probabilities using the approximate CCSD(T) Gibbs energies.
\fref{fig:phasediagram} depicts the the obtained contour lines for several reactions via different transition states with a probability of 0.02.
The choice of the probability introduces ambiguity but
we will mostly discuss trends and relative changes of these contour lines that
are not affected significantly by this choice.
We first discuss the contour lines for the transitions from low- to high-density phases as shown by
the dashed lines that approach the equilibrium phase boundary in the limit of higher temperatures
and turn to very large pressures in the limit of low temperatures.
From these dashed contour lines we can conclude that the ordering of the considered transition states
does not change in the investigated temperature and pressure range.
As such the transitions from the low- to high-density phases are always expected to proceed via
the \pcTS.
Furthermore the shape of these lines indicates that the activation of this transition depends strongly on
temperature and pressure. At low temperatures, high pressure is needed to drive this transition, whereas
significantly lower pressure suffices at higher temperatures.
We have also considered the transitions for the \hD\ to \cD\ and \wBN\ to \cBN\ phases.
The corresponding contours are depicted by the green almost vertical dashed lines, indicating
that the activation of the transitions is mostly temperature dependent.
We note that comparing the calculated green contour curves between carbon and boron nitride reveals that
they are shifted with respect to each other by about \SI{800}{\K}.
This observation is in agreement with experiment and reflects the fact that the barrier heights
in carbon are approximately \SI{180}{\meV\per\atom} larger than in boron nitride.
The shape of these contours are comparable with experimental findings
\cite{CorriganTJoCP1975,EremetsPRB1998,SolozhenkoHPR1995}.
In experiment the shape of these contour lines can be estimated by probing phase transitions
with and without catalysts for a range of temperatures and pressures.
In the transition from the wurtzite to the cubic phase,
bonds have to be broken to correct for stacking faults and twins in the crystal by
rearranging all boat to chair conformations.
This process requires
a minimum temperature to overcome the bond energy and drive the transition. 
We note that the vertical line in the experimental phase diagram of BN is at \SI{1500}{\K} and that
of carbon at \SI{2000}{\K}~\cite{BundyC1996,SolozhenkoHPR1995}.
The contour lines reveal an important problem in the synthesis of cubic diamond-like phases. 
When starting from a graphite-like phase a diamond-like phase can be obtained by applying
a minimum pressure, which increases strongly at lower temperatures.
However, once the metastable wurtzite phase is created or stacking faults and twins are present the kinetics
is very different and much higher minimum temperatures
are required to transform into the cubic phase or heal stacking faults.
This explains the almost rectangular region in the experimental phase diagram for the cubic phase.

\subsection{Revision of observed experimental phase transitions}\label{sec:revision}

\begin{figure*}[!ht]
  \centering
  \subfloat[\rBN]{\tikzmark{r}{\includegraphics[width=\figw]{BN-FC-1-G-ABC-POSCAR.jpg}\label{fig:rBN}}}\figs
  \subfloat[\BNAB]{\tikzmark{AB}{\includegraphics[width=\figw]{BN-T-1-G-AB-POSCAR-8.jpg}\label{fig:TG}}}\figs
  \subfloat[\hBN]{\tikzmark{h}{\includegraphics[angle=0,width=\figw]{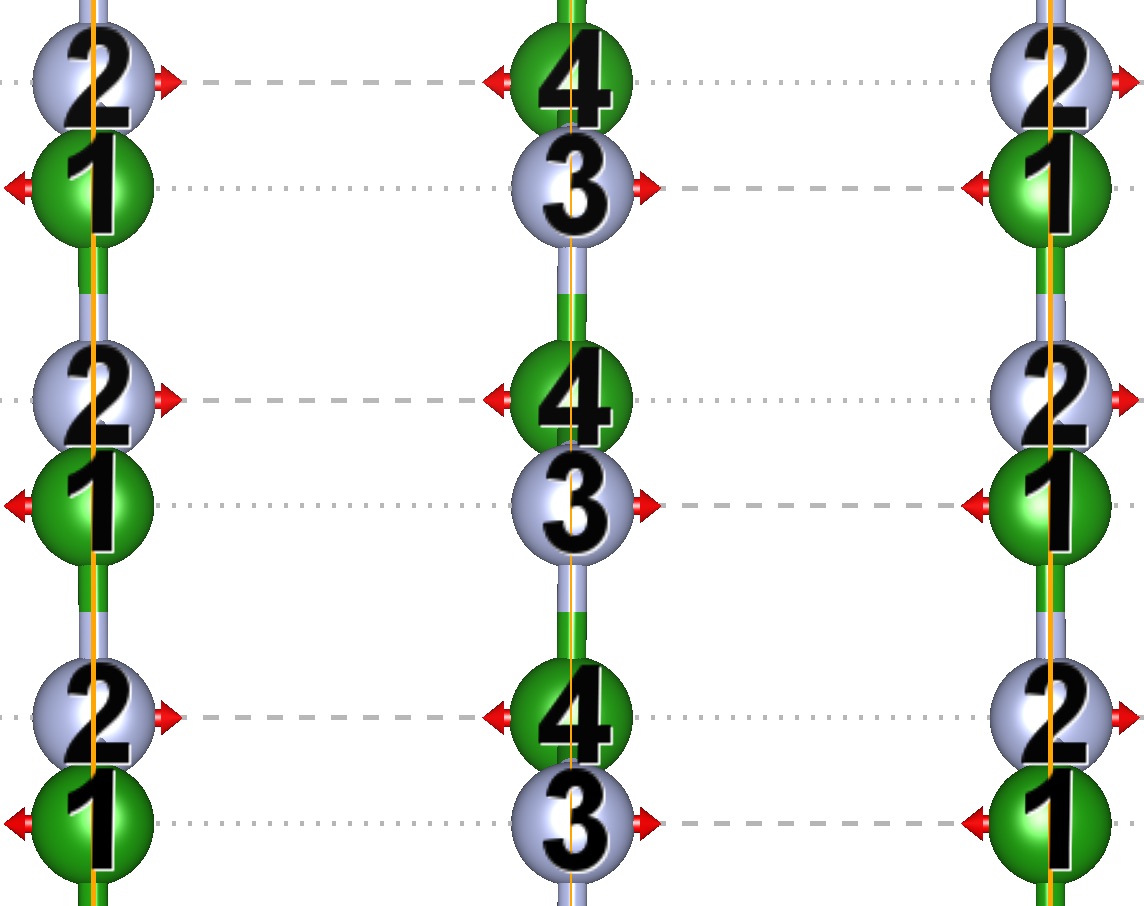}\label{fig:hBN}}}\figs
  \subfloat[\wBN]{\tikzmark{wc}{\includegraphics[width=\figw]{BN-1-D-H-D-C-POSCAR.jpg}\label{fig:wBNc}}}\\
  \subfloat[\pcTS]{\tikzmark{pc}{\includegraphics[width=\figw]{BN-FC-2-D-C-G-ABC-POSCAR.jpg}\label{fig:pcTS}}}\figs
  \subfloat[\bwTS]{\tikzmark{bw}{\includegraphics[width=\figw]{BN-T-2-D-H-G-AB-POSCAR.jpg}\label{fig:bwTS}}}\figs
  \subfloat[\pwTS]{\tikzmark{pw}{\includegraphics[angle=0,width=\figw]{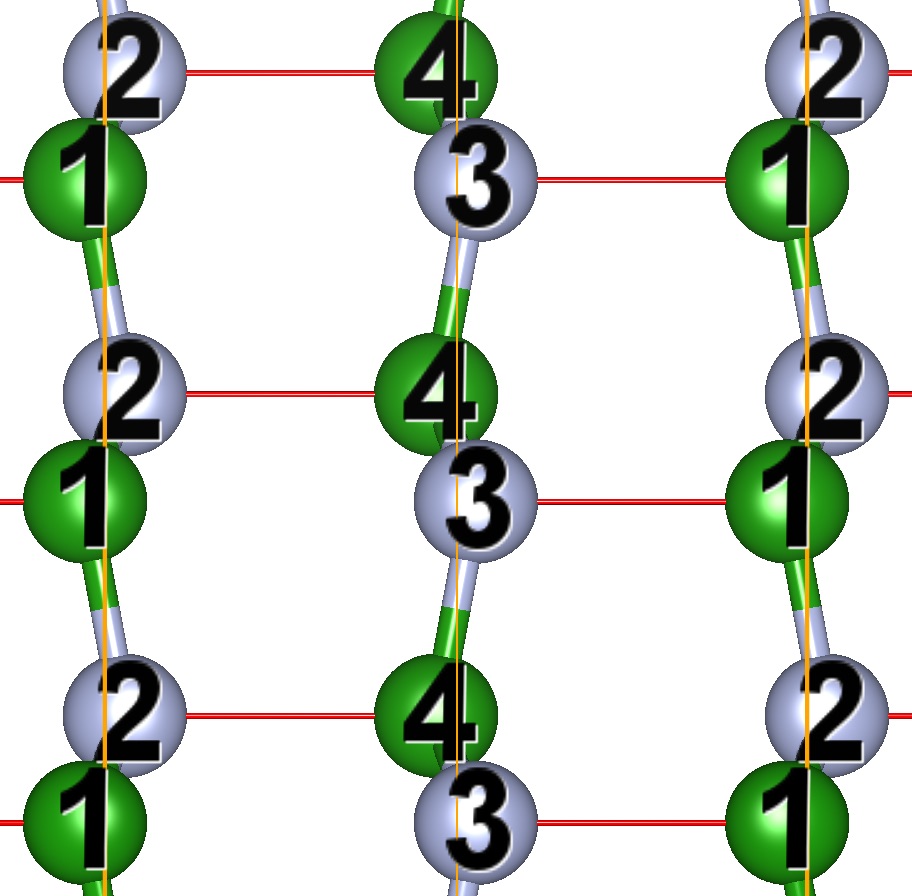}\label{fig:pwTS}}}\figs
  \subfloat[\lpcTS]{\tikzmark{lpc}{\includegraphics[width=\figw]{BN-2-D-H-D-C-POSCAR.jpg}\label{fig:lpcTS}}}\\
  \subfloat[\cBN]{\tikzmark{c}{\includegraphics[width=\figw]{BN-FC-3-D-C-POSCAR.jpg}\label{fig:cBN}}}\figs
  \subfloat[\wBN]{\tikzmark{w1}{\includegraphics[width=\figw]{BN-T-3-D-H-POSCAR-8.jpg}\label{fig:TD}}}\figs
  \subfloat[\wBN]{\tikzmark{w2}{\includegraphics[angle=0,width=\figw]{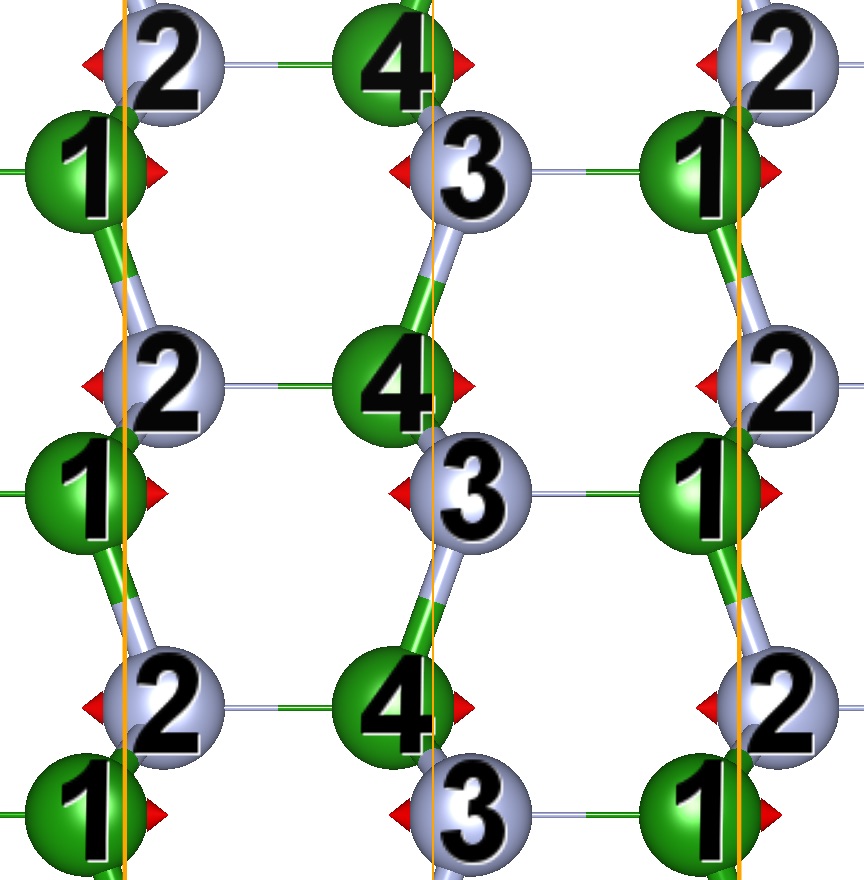}\label{fig:wBN}}}\figs
  \subfloat[\cBN]{\tikzmark{cw}{\includegraphics[width=\figw]{BN-3-D-H-D-C-POSCAR.jpg}\label{fig:cBNc}}}\\

  \ConnectC{r.south}{c.north}
  \ConnectC{AB.south}{w1.north}
  \ConnectC{wc.south}{cw.north}
  \ConnectD{r.south}{c.north}{green}
  \ConnectBN{h.south}{w2.north}
  \ConnectBN{wc.south}{cw.north}
  \Connectbow{r.south east}{w1.north}{30}{left}{green}{3cm}
  \Connectbow{r.south east}{h.south}{70}{right}{green}{5cm}
  \Connectbow{AB.south east}{cw.north}{40}{left}{black}{4cm}
 
  \caption{\label{fig:TSsum}Structures for different phase transition pathways:
  \protect\subref{fig:rBN} \ce{->[\text{\protect\subref{fig:pcTS}}\pcTS]} \protect\subref{fig:cBN}
  (BN: \cite{WentzcovitchPRB1988}; C: \cite{FahyPRB1986}),
  \protect\subref{fig:TG} \ce{->[\text{\protect\subref{fig:bwTS}}\bwTS]} \protect\subref{fig:TD}
  (BN: \cite{KurdyumovDaRM1996} for \rBN $\rightarrow$ \wBN; C: \cite{TateyamaPRB1996}),
  \protect\subref{fig:hBN} \ce{->[\text{\protect\subref{fig:pwTS}}\pwTS]} \protect\subref{fig:wBN}
  (BN: \cite{WentzcovitchPRB1988}; C: \cite{FahyPRB1987}),
  \protect\subref{fig:wBNc}~\ce{->[\text{\protect\subref{fig:lpcTS}}\lpcTS]} \protect\subref{fig:cBNc}
  (C: \cite{TateyamaPRB1996}).
  Red arrows indicate the atomic displacements and support together with the atom numbers the assignment during the phase transition.
  Dotted lines show new bonds to be formed and red lines represent strong interaction during the transition state.
  Black (carbon) and green (BN) arrows represents possible transitions (See text for further information).
  The crystal structures for C can easily be derived from BN, by substituting all B- and N-atoms with C-atoms.
  This makes \rBN, \BNAB, \hBN, \cBN\ and \wBN\ equivalent to \GABC, \GAB, \GAA, \cD\ and \hD, respectively.
  }
\end{figure*}

We now summarize experimental findings of observed phase transitions in more detail.
As an overview all ground and transition states are depicted in \fref{fig:TSsum}.
The first row shows the graphite-like phases (except for \wBN\ in the last column),
the transition states are in the middle row and the high-density phases in the bottom row.
The green and black arrows correspond to the experimentally observed transitions of boron nitride
and carbon, respectively.
A detailed discussion is provided in the following subsections. 

\subsubsection{Transitions between low and high density carbon phases }

Since graphite occurs in the AB stacking there is no direct transition pathway to \cD\ without
introducing shear stress.
This is illustrated by the yellow lines of \fref{fig:TG} and \ref{fig:cBNc}, which are not parallel to
each other.
Shear stress can be reduced by twin formation or stacking faults parallel to (111)$\lt{c}$, as can be
observed by high resolution TEM \cite{KulnitskiyACSB2013}. 
The stacking faults can be interpreted as \hD\ reflections \cite{NemethNC2014}. 
Another way of reducing shear stress is to bend the layers by applying pressure perpendicular
to the c-axis of graphite, inducing a so-called ``wave-like buckling and slipping`` mechanism \cite{XieSR2014}.
This could change the stacking order from AB to ABC and would make the \pcTS\ accessible,
reducing the activation energy since the \pcTS\ is the lowest transition state.

Experimental observations of the transformation from \GAB\ to \cD\ suggest a structural relationship
with plane (001)$\lt{G} \parallel$ (112)$\lt{c}$ and
direction [10$\overline{1}$0]$\lt{G} \parallel$ [111]$\lt{c}$ \cite{WheelerMRB1975}. 
The same orientation has been found in an MD simulation \cite{ScandoloPRL1995}.
This orientation fits exactly to the combination of
\fref{fig:TG} (\GAB) \ce{->[\text{\ref{fig:bwTS}}]} \ref{fig:TD}/\ref{fig:wBNc} \ce{->[\text{\ref{fig:lpcTS}}]} \ref{fig:cBNc} (\cD).
If graphite is compressed to over \SI{20}{\GPa} at room temperature the resistance increases, especially
perpendicular to the c-axis and returns  to its original value after
pressure is released \cite{BundyTJoCP1967,UTSUMIPotJASB1991}.
For temperatures between \SI{800}{\celsius} and \SI{1600}{\celsius} a
mixture of \hD\ and \cD\ is formed \cite{YoshiasaJJoAP2003}.
With a minimum pressure of \SI{12}{\GPa} and temperatures beyond \SI{1800}{\celsius} the portion of \hD\ decreases
and only \cD\ is left \cite{BundyC1996}.
Beyond \SI{1800}{\celsius} the temperature is high enough to reduce shear stress and to break bonds to heal
out stacking faults.
This is in good agreement with the calculated transformation probability contour lines depicted
in \fref{fig:phasediagram}.
\hD\ can be considered an intermediate structure and can be left out in the
conversion from  \GAB\ to \cD\ by slightly different
carbon displacements \GAB\ \ce{->[\lpcTS\ (\text{\fref{fig:lpcTS}})]} \cD. 
The \bwTS\ is higher in energy than the \lpcTS\ and the latter one would be preferred.
This means the \hD\ would not be created from the energetic point of view.
However, the latter mechanism is only probable if one accounts for the induced shear stress in some manner.

\GAB\ (\fref{fig:TG}) can transform into \hD\ without introducing shear stress.
The structure relation is the one via \bwTS\ (\fref{fig:bwTS}) as proposed
by Bundy and Kasper \cite{BundyTJoCP1967} and confirmed
many times \cite{WheelerMRB1975,YagiPRB1992,YoshiasaJJoAP2003}.

The activation energies for the carbon system are higher than for boron nitride and therefore
require higher pressures and larger temperatures to observe.
For the catalytic transition at the phase boundary the activation energy in solution is
about \SIrange{1.3}{1.7}{\eV} \cite{BundyC1996}.
However, this is not comparable with the calculated solid-solid phase transitions from this work
since the carbon dissolves within the liquid catalyst.

\subsubsection{Transitions between low and high density BN phases }

For the boron nitride system the direct transitions via \pcTS\ and \pwTS\ are the most probable
because the stable graphite-like structures exhibit an ABC (\rBN) and AA' (\hBN) stacking
and these transition states have the lowest activation energies of those calculated in the present work.
Experimentally direct conversion of \hBN $\rightarrow$ \wBN\ and
\rBN $\rightarrow$ \cBN\ have been observed.
Experimental observations of structural orientation relationships of initial and final states
can be used to infer which transition states are possible.
Experimentally the transitions \rBN $\rightarrow$ \cBN\ (\fref{fig:rBN} \ce{->[\text{\ref{fig:pcTS}}]}
\ref{fig:cBN} with plane (001)$\lt{r} \parallel$ (111)$\lt{c}$ and direction [11$\overline{2}$0]$\lt{r}
\parallel$ [110]$\lt{c}$) and \hBN $\rightarrow$ \wBN\ (\fref{fig:hBN} \ce{->[\text{\ref{fig:pwTS}}]}
\ref{fig:wBN} with plane (001)$\lt{h} \parallel$ (001)$\lt{w}$ and direction [10$\overline{1}$0]$\lt{h}
\parallel$ [10$\overline{1}$0]$\lt{w}$) has been observed during shock wave synthesis \cite{SatoJotACS1982}.
Due to the different stacking sequence of \GAB\ and \hBN\ the structure
relationship to \hD\ and \wBN\ is different \cite{TaniguchiAPL1997}.
The 90$^{\circ}$ rotation can be seen by comparing \GAB\
(\fref{fig:TG}) \ce{->[\bwTS\ (\text{\fref{fig:bwTS}})]} \hD\ (\fref{fig:TD})
with \hBN\ (\fref{fig:hBN}) \ce{->[\pwTS\ (\text{\fref{fig:pwTS}})]} \wBN\ (\fref{fig:wBN}).
The minimum pressure and temperature conditions for the direct transitions
of \rBN\ \ce{->[\text{\pcTS}]} \cBN\ at about \SI{1200}{\celsius} and more
than \SI{8}{\GPa} \cite{TaniguchiDaRM1997} are slightly lower than
for \hBN\ \ce{->[\text{\pwTS}]} \wBN\ with about \SI{1400}{\celsius} and more
than \SI{8.5}{\GPa} \cite{CorriganTJoCP1975}.
This agrees with the higher activation energy of \pwTS\ compared to \pcTS\ (rate curves in \fref{fig:phasediagramBN}).
Above \SI{13}{\GPa} \hBN\ transforms partially to \wBN\ at room temperature, and completely at
temperatures exceeding \SIrange{800}{1000}{\celsius} \cite{CorriganTJoCP1975}.
This implies that for the transition from low to high density phases the pressure is crucial,
the higher the pressure the less temperature is needed, which is reflected in the transition
probability contour lines shown in \fref{fig:phasediagramBN}.
It has also been reported that \rBN\ can transform into \wBN\ above \SI{8}{\GPa} and at
low temperatures (\SIrange{25}{400}{\celsius}) via \bwTS\ \cite{KurdyumovDaRM1996,TaniguchiDaRM1997}. 
However, these samples of \wBN\ (formed from \rBN) contain a lot of stacking faults.
We find that \bwTS\ is the least favorable transition state and it is more likely that \cBN\ is
formed with stacking faults that create stacks of \wBN\ due to stress and low repairing ability
at low temperatures.

\cBN\ can not only be created from \rBN\ (via \pcTS\ \fref{fig:pcTS}), but also
from \wBN\ (via \lpcTS\ \fref{fig:lpcTS}).
Therefore a transition from \hBN\ to \cBN\ could proceed via \wBN\ as an intermediate
structure \cite{Gmelin1988,BritunJoMS1993}. 
However, there is no experimental proof for such a direct conversion from \hBN\ to \cBN.
At elevated pressure \hBN\ first converts into \wBN.
However, only at a minimum pressure and temperature of about \SI{8}{\GPa} and \SI{1400}{\celsius} \wBN\
can start converting into \cBN\ but lacks a clear equilibrium phase boundary,
indicating that \wBN\ is a meta-stable phase \cite{OnoderaTJoCP1981}.
The resulting \cBN\ is not pure and contains fractions of \wBN.
Only at very high pressures and temperatures  ($\sim$\SI{20}{\GPa} and \SI{2300}{\celsius}) the resulting
product is pure \cBN\ \cite{NagakuboAPL2013}.
As such for the transition from \wBN\ to \cBN\ temperature is the limiting factor.
The transition starts at \SI{1500}{\celsius} and more than \SI{2000}{\celsius} is needed to
complete \cite{CorriganTJoCP1975,NagakuboAPL2013}. 
Increasing the pressure rather increases the transition temperature and our probability curve
confirms that (\fref{fig:phasediagramBN}).

\wBN\ transforms into \hBN\ at zero pressure and above \SI{1300}{\celsius} \cite{Gmelin1988}. 
With increasing grain size of \cBN\ powder the onset temperature for the transition to \hBN\ increased
between \SIrange{900}{1500}{\celsius}. 
Its transition is usually accompanied by a significant formation of cracking, which could be a sign for
no direct transition from \cBN\ to \hBN\ \cite{SachdevDaRM1997}.

The experimental values for the activation energy of the \wBN\ $\rightarrow$ \cBN\ reaction span
a wide range of \SIrange{0.96}{3.4}{\eV} depending on the reaction conditions \cite{Gmelin1988,Gmelin1991}.
The value for shock wave synthesis is even higher: \SI{8.7}{\eV\per\atom} \cite{CorriganTJoCP1975}.
Since all these values for the \wBN\ $\rightarrow$ \cBN\ reaction are obtained from non equilibrium
conditions, they can not be compared with the calculated ones.
Obtained activation energies are influenced by kinetic effects, which arise from grain
size, defects or other structural distortions \cite{OnoderaTJoCP1981}.
Therefore these values are rather upper bounds than real activation energies.
If these were real activation energies, the backward reaction would have a similar activation energy.
However, this is not true as can be seen by comparing with the \wBN\ $\rightarrow$ \hBN\ reaction
with an experimental activation energy of \SI{0.22}{\eV\per\atom}~\cite{WillsIJoHTC1985}. 
The latter value is comparable with the calculated one at the same experimental conditions
($\sim$ \SI{0.17}{\eV\per\atom} at \SI{900}{\celsius}). 

\subsubsection{Transition between low density graphite-like BN phases}

The \rBN\ has an ABC stacking and could be converted to AB or AA trough translation because the layers are only shifted towards each other (\fref{fig:LAB} and \ref{fig:LABC}).
\hBN\ has an AA' stacking, where each layer is rotated by 60$^\circ$ towards their neighboring layers and a translation would never create an AA, AB or ABC stacking (\fref{fig:layer}).
To reach \bwTS\ and \pwTS\ pressure has to be applied.
The transformation from \rBN\ $\rightarrow$  \hBN\ occurs only at elevated pressure and by applying shear stress \cite{TaniguchiDaRM1997}, which could be a sign, that the layers have first to be changed into the AD stacking order and buckle before they can transform into each other.
One possible transition path could be similar to \fref{fig:rBN} \ce{->[\text{\ref{fig:TG}}]} \ref{fig:bwTS} by changing the layer sequence from ABC to AD and result in the \bwTS\ or close to it.
It can switch to the \pwTS\ (\ref{fig:pwTS}) by changing some bond lengths.
This changes the orientation of the c-axis and the new c-axis has an AA' stacking order and after de-puckering and elongation along this new c-axis the \hBN\ structure is formed.

\subsection{Hexagonal diamond (Lonsdaleite)}\label{sec:lonsdaleite}

While the synthesis of high quality wurtzite boron nitride crystals is possible,
the existence of the corresponding carbon polymorph known as lonsdaleite has recently been called into question~\cite{NemethNC2014}.
A number of studies have investigated the wurtzite structure of carbon and shown
the main XRD peaks of the cubic phase are also part of the XRD spectra of the wurtzite phase,
making it difficult to distinguish these phases.
Ignoring the relative intensities of the peaks maybe due to textures effects, makes it
impossible to determine the exact amount of the cubic phase within the wurtzite phase.
The latter problem occurred in older publications about diamond, where
the detected lines only have been published without the XRD spectra~\cite{BundyTJoCP1967,YagiPRB1992}.
The largest peak of the \hD\ spectra which is not part of the \cD\ spectra has a \textit{d}-spacing
of \SI{2.18}{\text{\AA}} and is just a shoulder of the main peak with a \textit{d}-spacing
of \SI{2.06}{\text{\AA}}.
In a recent publication of XRD spectra for a natural and a synthetic sample the \hD\ peaks are
just shoulders and the sample is mainly \cD\ \cite{NemethNC2014}.
The authors point out that the peaks are due to stacking faults of basal planes and twinning, which are
supported by STEM images.
The defects create new planes with different \textit{d}-spacing compared to cubic single crystals and have
the same spacing like the ones in \hD\ because these planes are in the wurtzite structure. 
Therefore the amount of \hD\ can be assigned to the amount of defects.
Transforming \GAB~\ce{->[\pcTS]}~\cD\ implies to produce shear stress due to changing the stacking order
from AB to ABC, as already pointed out by Tateyama \cite{TateyamaPRB1996}.
By forming a (111) twin the stacking order inverts and the stress is reduced.
The formation of one type of twins can be seen in an MD simulation \cite{ScandoloPRL1995}.
The twin planes where the stacking order inverts corresponds to the \hD\ structure.
This is a clear example why \hD\ can not by synthesized as a single crystal easily.
Yoshiasa \etal\ observed a higher ratio of \hD\ to \cD\ for X-ray diffraction profiles perpendicular
to the c-axis of the parent graphite \cite{YoshiasaJJoAP2003}.
This supports that \hD\ stacking is not produced by the \pwTS, but via \bwTS\ because of
the orientation of the c-axis.
The pressure and temperature region observed by Bundy and Kasper \cite{BundyTJoCP1967} for the
formation of \hD\ has been investigated by other groups and not all were able to synthesize
a detectable amount of \hD\ \cite{EndoPRB1994}, which was attributed it to the different experimental conditions.

The same type of defects as described above were also observed in the BN system \cite{KurdyumovDaRM1996}.
However, in contrast to  \hD, \wBN\ can be synthesized relatively pure \cite{NagakuboAPL2013}.
We attribute this to the existence of a stable graphite-like (\hBN) structure together with a relatively low
transition state (\pwTS) that forms directly the wurtzite phase without inducing shear stress.
In the carbon system the graphite-like phase with the same stacking order as \hD\ (\GAA) does not exist.
\GAB\ needs to shift individual sheets to reach the \bwTS\ or \lpcTS. This is very unlikely to happen and would
also require larger temperatures.
Since \hD\ and \wBN\ are meta-stable structures an increase in temperature and pressure will always
lead to a transformation into the thermodynamically stable \cD\ and \cBN\ structures.

\section{Conclusion}

In this work we have investigated (meta-)stable boron nitride as well as carbon
allotropes for a range of pressures and temperatures.
Furthermore corresponding concerted transition pathways have been explored.
The calculations were performed using a selection of approximate exchange and correlation
density functionals and quantum chemical wavefunction based theories including the
coupled cluster method.
A comparsion between the theoretical and experimental findings reveals
that highly accurate predictions for equilibrium phase boundaries
constitute a true challenge for state of the art electronic structure theories.

We have investigated the energy differences between low- and high-density phases of
carbon and boron nitride.
Due to the variation in the results obtained using LDA, GGA, mGGA and hybrid functionals a firm
conclusion and accurate estimate of the energy differences can not be achieved.
Furthermore the explicit inclusion of van der Waals interactions on the level of MBD is found to
be significant and might change the order of the predicted stability depending on the employed parent XC functional.
We stress that considering other approximations to the van der Waals interactions or additional XC functionals
would not allow for achieving more reliable results.
On the other hand we find that quantum chemical wavefunction based theories allow for a systematic improvability
of the obtained results. HF, MP2, CCSD and CCSD(T) methods yield an oscillating but convergent estimate
of the calculated energy differences. We note in passing that such a systematic behaviour was recently also
reported for calculated transition pressures in  LiH crystals~\cite{doi:10.1063/1.4928645}.
The CCSD(T) method predicts that the corresponding low- and high-density phases of boron nitride
as well as carbon are degenerate to about \SIrange{10}{20}{\meV\per\atom} including ZPVEs.
We stress that the remaining uncertainty of coupled
cluster theory results is dominated by finite size effects that can possibly be further reduced in future
by studying larger systems. We also note that finite size errors are significantly larger for results obtained
using second-order M\o ller-Plesset perturbation theory (MP2) in particular for small gap systems,
where MP2 is considered less accurate.
The present coupled cluster theory results for the energy difference between carbon diamond and graphite are
in agreement with experimental measurements and quantum monte carlo calculations from literature to
within about \SIrange{20}{30}{\meV\per\atom}, which corresponds to the accuracy that is typically ascribed to 
CCSD(T) theory.
The same conclusion can not be drawn for boron nitride due to a larger spread in the available experimental findings.
However, we hope that this work will motivate further calculations using 
quantum monte carlo methods and experimental studies to help providing
more accurate estimates of the corresponding equilibrium phase boundaries.

The obtained coupled cluster theory results
for the activation barrier heights in the graphitic to diamond-like transitions of boron nitride as well as carbon
also allow for
benchmarking different levels of approximate exchange and correlation density functionals.
We conclude that the accuracy of the employed LDA, GGA and mGGA functionals follows roughly the same
trends as for activation barrier heights in molecular gas phase reactions: LDA, GGA and mGGA functionals
underestimate the barrier heights if the effect of van der Waals interactions is taken into account.
Furthermore the results for hybrid functionals indicate a strong dependence on the choice of parametrization.
We find that PBE0 and HSE06 yield significantly more accurate results than B3LYP, confirming
previous findings for a wide range of solids~\cite{doi:10.1063/1.2747249}.
Furthermore we note that the investigated transition states are not very strongly correlated
as indicated by the good agreement of a few \SI{10}{\meV\per\atom} between CCSD and CCSD(T) theory.
The observed finite size effects are larger for the predicted coupled cluster barrier heights
than for the energy differences of the (meta-)stable allotropes.
An important conclusion for the investigated transition states is that their ordering and
relative stabilities is mostly independent from the employed electronic structure theory.
All employed theories predict unequivocally that the puckering mechanism as present in the \pcTS\ is
energetically the most favorable transition mechanism for boron nitride as well as carbon.

The prediction of pressure-temperature phase diagrams requires the calculation of Gibbs energies.
We have shown that approximating the CCSD(T) Gibbs energy using
the CCSD(T) energies of the (meta-)stable and transition states only
and the LDA for its temperature and pressure dependence yields reliable pressure-temperature phase diagrams.
The obtained phase boundaries agree with experimental results of carbon to within about one GPa at 
temperatures below 2000~K. In the case of boron nitride we find a similarly good agreement with a
recently obtained experimental result of Fukunaga et. al. in Ref.~\cite{FukunagaDaRM2000}.
Furthermore we have provided estimates of
approximate phase transition probabilities in a similar manner.
The calculated phase transition probabilities confirm trends in the measured pressure and
temperature dependence of experimentally observed phase transitions.

Finally we have addressed the conversion of graphite to hexagonal diamond also known as
lonsdaleite using the obtained results for transition and (meta-)stable states.
In the context of the present work it is reasonable to ask the question:
why can the wurtzite phase of boron nitride be synthesized as an almost pure powder whereas
the existence of single crystals of
lonsdaleite is still under debate? We conclude that the puckering mechanism for the corresponding phase transitions is always
the most probable due to its energetically more favorable transition state.
However, the stacking of the parent graphitic phase that is put under pressure
has a significant influence on the kinetics of the phase transition.
We note that cubic diamond and wurtzite structures exhibit an ABC and AA' stacking, respectively.
Experimentally \wBN\ is formed only when applying pressure to \hBN, which exhibits
also an AA' stacking. In the case of carbon the corresponding \GAA\ phase is not stable, making
a transformation from \GAB\ or \GABC\ to lonsdaleite only possible by introducing stacking faults or similar
defects. This conclusion is in agreement with recent experimental work.

\begin{acknowledgments} 
This project has received funding from the European Research Council (ERC) under the
European Union’s Horizon 2020 research and innovation program (grant agreement No 715594).
The computational results presented have been achieved in part using the Vienna Scientific Cluster (VSC).
Helpful discussions with Ali Alavi are gratefully acknowledged
\end{acknowledgments}

\bibliographystyle{apsrev4-1}
\bibliography{Quellen,cite_grueneis}

\newpage

\end{document}